\pdfoutput=1

\documentclass[sigconf]{acmart}
\usepackage{float}
\usepackage{listings}
\usepackage{verbatim}
\usepackage{supertabular}

\usepackage{tabularray}
\usepackage{caption}
\usepackage{multicol}
\usepackage{outlines}

\usepackage{subcaption}
\usepackage{array}
\usepackage[normalem]{ulem}
\usepackage{todonotes}
\usepackage{soul}
\usepackage{xspace}
\usepackage{xcolor}
\usepackage{longtable}
\usepackage{tabularx}
\usepackage{graphicx}
\usepackage{enumitem,kantlipsum}

\newcommand{\nickname}{LogoMotion\xspace}
\definecolor{lightgray}{gray}{0.6}

\newcommand{\revAdd}[1]{\textcolor{black}{#1}} % make added text (argument #1) blue
\ifdefined\final
\else
\newcommand{\final}{0}
\fi

\newcommand{\vivian}[1]{{\color{purple} Vivian: #1}}
\newcommand{\lydia}[1]{{\color{purple} Lydia: #1}}
\newcommand{\liyi}[1]{{\color{orange} Li-Yi: #1}}
\newcommand{\rubaiat}[1]{{\color{blue} Rubaiat: #1}}
\newcommand{\tim}[1]{{\color{green} Tim: #1}}

\ifthenelse{\equal{\final}{1}}
{
\renewcommand{\vivian}[1]{}
\renewcommand{\lydia}[1]{}
\renewcommand{\liyi}[1]{}
\renewcommand{\rubaiat}[1]{}
\renewcommand{\tim}[1]{}
}
{}

\newcolumntype{C}[1]{>{\centering\arraybackslash}p{#1}}

\usepackage{listings}
\usepackage{xcolor}

\definecolor{codegreen}{rgb}{0,0.6,0}
\definecolor{codegray}{rgb}{1,1,1}
\definecolor{codepurple}{rgb}{0.58,0,0.82}
\definecolor{backcolour}{rgb}{0.8,0.8,0.8} % Black background
\definecolor{textcolour}{rgb}{0,0,0} % White text

\AtBeginDocument{%
  \providecommand\BibTeX{{%
    \normalfont B\kern-0.5em{\scshape i\kern-0.25em b}\kern-0.8em\TeX}}}
\setcopyright{acmcopyright}
\copyrightyear{2018}
\acmYear{2018}
\acmDOI{10.1145/1122445.1122456}

\acmConference[Woodstock '18]{Woodstock '18: ACM Symposium on Neural
  Gaze Detection}{June 03--05, 2018}{Woodstock, NY}
\acmBooktitle{Woodstock '18: ACM Symposium on Neural Gaze Detection,
  June 03--05, 2018, Woodstock, NY}
\acmPrice{15.00}
\acmISBN{978-1-4503-XXXX-X/18/06}

\copyrightyear{2025}
\acmYear{2025}
\setcopyright{acmlicensed}\acmConference[CHI '25]{CHI Conference on Human Factors in Computing Systems}{April 26-May 1, 2025}{Yokohama, Japan}
\acmBooktitle{CHI Conference on Human Factors in Computing Systems (CHI '25), April 26-May 1, 2025, Yokohama, Japan}
\acmDOI{10.1145/3706598.3714155}
\acmISBN{979-8-4007-1394-1/25/04}

\begin{document}

\title{\nickname: Visually-Grounded Code Synthesis for Creating and Editing Animation} %


\begin{abstract}

Creating animation takes time, effort, and technical expertise. To help novices with animation, we present LogoMotion, an AI code generation approach that helps users create semantically meaningful animation for logos. LogoMotion automatically generates animation code with a method called visually-grounded code synthesis and program repair. This method performs visual analysis, instantiates a design concept, and conducts visual checking to generate animation code. LogoMotion provides novices with code-connected AI editing widgets that help them edit the motion, grouping, and timing of their animation. In a comparison study on 276 animations, LogoMotion was found to produce more content-aware animation than an industry-leading tool. In a user evaluation (n=16) comparing against a prompt-only baseline, these code-connected widgets helped users edit animations with control, iteration, and creative expression.

\end{abstract}





\begin{teaserfigure}
\centering
  \includegraphics[width=\textwidth]{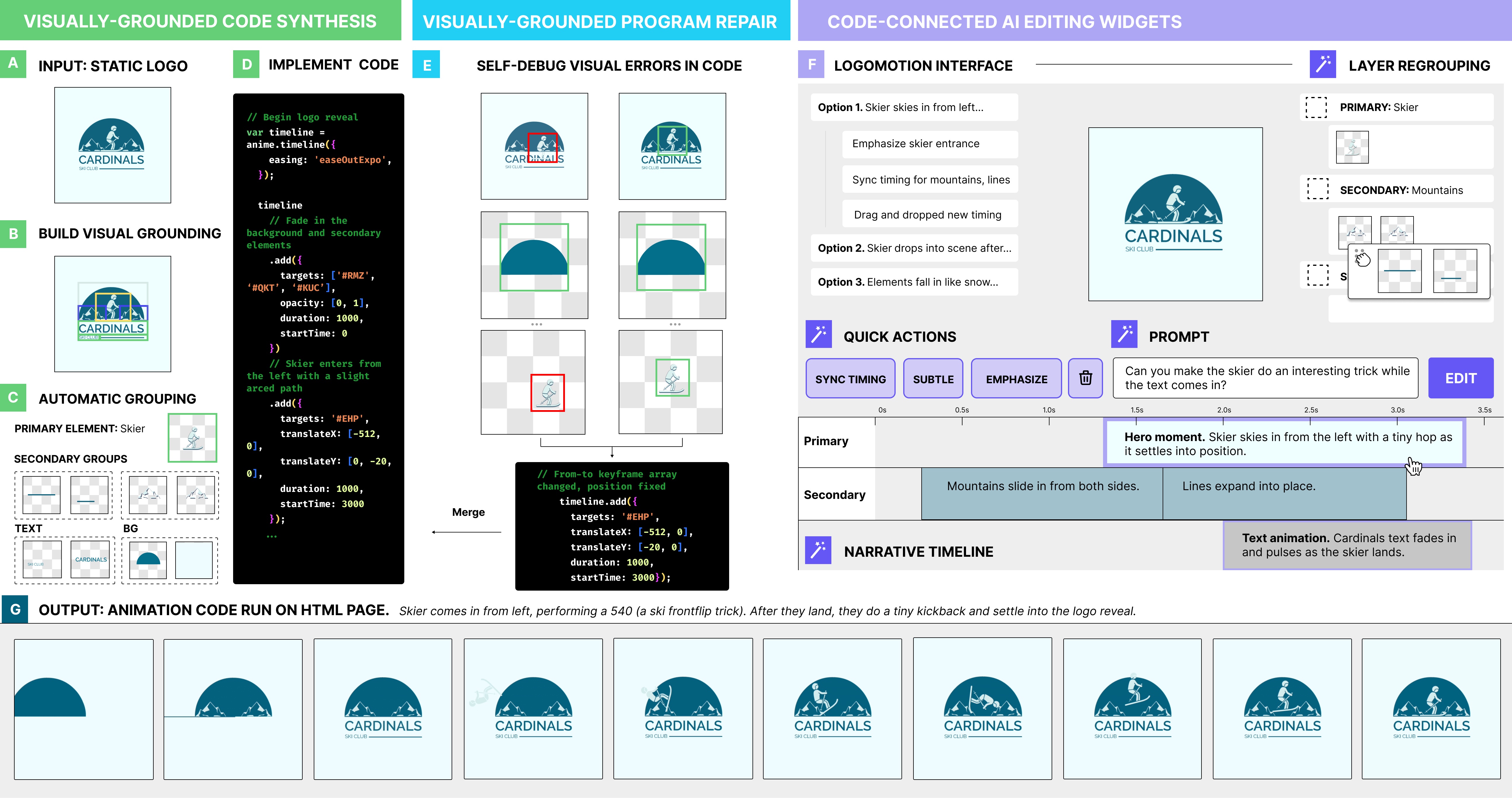}
  \caption{  \textbf{\nickname} is an AI code generation approach that helps users create semantically meaningful animation for logos. \nickname automatically generates animation code for a logo design using \textbf{visually-grounded code synthesis and program repair}. This method performs visual analysis, instantiates a design concept, and conducts visual checking to generate animation code. Novices can use \textbf{code-connected AI editing widgets} to make targeted edits to the motion, grouping, and timing of their animation through familiar GUI controls.}
  \Description{The left side of the image shows a static layered logo with a skier and design elements (mountains, lines) on it. The text says CARDINALS, SKI CLUB. The first column is VISUALLY GROUNDED CODE SYNTHESIS. A) INPUT: STATIC LOGO. B) BUILD VISUAL GROUNDING. C) AUTOMATIC GROUPING. Underneath C, layers are shown in AI-assigned groups. D) shows code that is implemented using this visual understanding. (2. VISUALLY-GROUNDED PROGRAM REPAIR), a mechanism for self-debugging visual errors is described (labeled E). Two images show a visual difference between the placement of the primary element at the last frame of the animation and where it should be in the target layout. This difference is shown isolated on individual layers. A code snippet is generated to fix the issue and an arrow saying Merge indicates how visual feedback from the canvas is reflected in code. The column next to it shows the code that is implemented. The third column DYNAMICALLY GENERATED ACTION PANEL FOR ANIMATION EDITING shows the LogoMotion interface (F), which has a layer panel for layer regrouping, a canvas, a prompt box, quick action (SYNC TIMING, SUBTLE, EMPHASIZE) and a narrative timeline with tracks: PRIMARY, SECONDARY, shown. At the bottom, a row of frames from the animation shows the output (ANIMATION CODE RUN ON HTML PAGE): Skier comes in from left, performing a 540 (a ski frontflip trick). After they land, they do a tiny kickback and settle into the logo reveal.}
  \label{fig:teaser}
\end{teaserfigure}

\author{Vivian Liu}
\email{vivian@cs.columbia.edu}
\affiliation{
  \institution{Columbia University}
  \city{New York}
  \state{NY}
  \country{USA}
}

\author{Rubaiat Habib Kazi}
\email{rhabib@adobe.com}
\affiliation{
  \institution{Adobe Research}
  \city{Seattle}
  \state{WA}
  \country{USA}
}

\author{Li-Yi Wei}
\email{liyiwei@acm.org}
\affiliation{
  \institution{Adobe Research}
  \city{San Jose}
  \state{CA}
  \country{USA}
}

\author{Matthew Fisher}
\email{matfishe@adobe.com}
\affiliation{
  \institution{Adobe Research}
  \city{San Francisco}
  \state{CA}
  \country{USA}
}

\author{Timothy Langlois}
\email{tlangloi@adobe.com}
\affiliation{
  \institution{Adobe Research}
  \city{Seattle}
  \state{WA}
  \country{USA}
}

\author{Seth Walker}
\email{swalker@adobe.com}
\affiliation{
  \institution{Adobe Research}
   \city{San Francisco}
  \state{CA}
  \country{USA}
}

\author{Lydia Chilton}
\email{chilton@cs.columbia.edu}
\affiliation{
  \institution{Columbia University}
    \city{New York}
  \state{NY}
  \country{USA}
  % \city{New York}
  % \state{New York}
}
\maketitle

\lstset{
    basicstyle=\footnotesize\ttfamily,
    % specifies that lines shouldn't overrun the margin
    breaklines=true
    % specified that, if possible, lines should be broken at whitespace
    breakatwhitespace=true
}

\section{Introduction}

Motion suggests life \cite{socialshapes}. Motion designers use animation to signal change, suggest relationships between objects, convey personality, and connect objects to the actions and activities they take on in the real world. They make cars drive, bumblebees fly, and stars twinkle, because people have visual expectations for how these objects should move. The dynamics of animation can further communicate meaning. A delivery truck driving in fast onto the screen can symbolize speedy service; the way a circle vertically moves can symbolize something as gentle as a sunrise or something as playful as a bouncing ball. Animation is delightful when it is semantically meaningful: when it ``reflects and reinforces the semantic relationships" between design elements \cite{ux_animation}.

% low-level interactions with keyframes and 
Currently, creating semantically meaningful animation is difficult for novices. Professional tools like After Effects allow this, but they can require technical expertise and fine-grained editing that can be time consuming. End-user animation tools such as Canva and Adobe Express are popular with novice designers and provide support for animation. However, their animation tools are largely based on animation presets and templates. For example, a novice could want to animate a logo for a Ski Club logo that has a skier, some mountains, lines, and text (as pictured in Fig. \ref{fig:teaser}). They can apply an animation preset that automatically animates the page and has everything "come in from the left" or "fade in". They could also edit the speed or intensity of the motion in parameters or apply other preset motion styles such as "energetic" or "professional". However, while templates and presets can create smooth, well-executed animation, they can be limited at creating motion that truly reinforces the logo's visual message (e.g. making a skier ski in and do a flip). 

Generative AI can move beyond rigid and rule-based systems like templates and presets to provide more flexible means of creation. It can generate animation code that is more complex and expressive than presets. Furthermore, AI has the multimodal capability to perform visual analysis and understand the semantics of an image, such as what each element depicts (a skier, a mountain) and what role each element plays on the canvas (foreground, background, groups). For example, an AI can visually analyze that the skier in the logo (Fig.~\ref{fig:teaser}-A) is the primary element and that there are a group of mountains. Using this analysis, it can come up with a design concept (``have the skier ski in from the left and do a flip into place, the mountains rise in, and the text enter as the skier lands") and implement code executing that concept. We call this \textbf{visually-grounded code synthesis}, a method that performs visual analysis and instantiates a design concept to guide the AI implementation of animation code . However, AI-generated code can have errors, and AI approaches should help users fix the errors they create. We introduce \textbf{visually-grounded program repair} to run AI-generated code, visually inspect the animation result for errors (``skier position is off"), and let the AI automatically debug itself. Together, visually-grounded code synthesis and repair generate an animation that is semantically meaningful to the logo design (Fig.~\ref{fig:teaser}-G).

Although this automatic logo animation approach often generates good results by itself, users should be able to edit and perfect it. Prompt-based editing of code has potential \cite{Tseng:24:Keyframer, Angert:23:Spellburst} but can lack the control necessary for editing animation. A prompt can regenerate the entire animation code, change parts the user likes, or introduce new errors as the AI tries to fulfill the user's desired edit. 
% Editing through PROMPTING is sometimes effective, but it's hard to specify low level actions like how to specially synchronous the timing of two elements. other systems EXPECT users to get fined grained edits, but editing the code. 
To offer users more control over the editing of generated code, we introduce
\textbf{code-connected AI editing widgets}. \nickname's editing UI contains familiar animation editing widgets such as a timeline, a layer panel, and quick actions ("more subtle", "faster", "synchronize timing"). When the user interacts with these widgets, \nickname translates the GUI interaction into targeted edits over the code corresponding to the motion, grouping, and timing of the animation.
% To enable this, the system has to connect the code to the widgets when it is created. 
% When the widgets (say the timeline) is created, the system looks at the AI generated code and CONNECTS THE CODE TO BUTTONS? 
% Then, when the user interacts with the widget, the code knows specifically what elements to update and with what operation. 
% Knowing exactly what code to touch is important so that what the user wants to edit is updated, but the rest of the code stays the same.
% this is how AI  gives users control over editing.
These code-connected AI editing widgets let users gain the expressiveness and controllability of a code representation while having a visual representation that lets them focus on the creative aspects of editing.

Our contributions are as follows:
\begin{outline}

\1  \textbf{Visually-Grounded Code Synthesis and Program Repair} automatically generates semantically meaningful animations for logo designs. This method performs visual analysis, provides a design concept, and applies visual checking over the canvas to generate animation code.

\1  \textbf{Code-Connected AI Editing Widgets} enable users to edit AI-generated code with widgets such as a narrative timeline, layer panel, and quick actions. These widgets bind GUI controls to targeted code edits, so users can edit the motion, grouping, and timing of their animation.

\1 Two technical evaluations showing that 1) in comparison study of 276 animations, visually-grounded code synthesis produces more semantically meaningful animations than an industry-leading animation tool, 2) visually-grounded program repair can solve 96\% of detected errors within four attempts.

\1 A user study (n=16) comparing with a prompt-only baseline showed that code-connected AI widgets helped novices edit with control, iteration, and creative expression.

\end{outline}

We conclude with a discussion of how \nickname's methods of combining code representations with visual understanding and code-connected widgets can apply to other design tasks beyond logo animation.

\section{Related Work}
\subsection{Program Synthesis}

Program synthesis, the formal name for code generation, is the idea that given a high-level specification of a problem, a search space of potential program solutions can be automatically searched to find a provably correct solution \cite{msr}. While program synthesis originated in the domain of formal methods and boolean SAT solvers, it has evolved greatly since the introduction of machine learning and large language models. The state of the art models for code generation include GPT-4, AlphaCode, CodeGEN, Code Llama, and Gemini \cite{gpt4, alphacode, codegen, codellama, gemini}. These models generally take in a natural language specification of the problem (e.g. docstrings), test cases, and examples of inputs and outputs. They have shown remarkable ability at being able to solve complex programming problems at the level of the average human programmer \cite{alphacode}. Prompting for code generation generally differs from traditional prompting interactions, because code has underlying abstract syntactic representations, while natural language prompts can be more declarative and focused on conceptual intent \cite{fiannaca23prompting}.  Generating code often involves intermediate representations such as scratchpads \cite{nye2021scratchpads}, chain-of-thought, and chain-of-code operations \cite{li2023chain, nl2spec}.

While code generation models have primarily been benchmarked on text-based programming problems (e.g. LeetCode problems), they have also been shown to capably handle visual tasks. ViperGPT demonstrated that a code generation model can be used to compose computer vision and logic module functions into code plans to answer visual queries \cite{vipergpt}. HCI systems have shown that code generation models can be integrated within creative workflows to provide interactive assistance \cite{Tseng:24:Keyframer, spellburst}. Spellburst demonstrated how LLMs can help end users explore creative coding through natural language prompts and operations that merge Javascript code scripts \cite{spellburst}. BlenderGPT allows users to use prompts to handle 3D tasks such as scene creation, shader generation, and rendering \cite{blendergpt}. While other works also build text representations of a visual output (e.g. a 3D scene hierarchy \cite{llmr} or JSON representation of UI \cite{automaticfeedback}), \nickname demonstrates how these text representations can be augmented with canvas understanding to visually ground code generation.

% As in these earlier works, code generation models often compose abstractions from libraries that were written to programmatically create visuals (bpy, CSS, p5.js) \cite{bpy, css_animation}

% \rubaiat{should we clarify what is the missing gap we are bridging with our method? how is our method new/different from the above ones?}

A recent direction within the program synthesis space has been the use of LLMs to inspect and edit the code they generate for program repair (also known as self-refinement or self-debugging) \cite{chen2023selfdebug}. Traditionally, program repair approaches primarily focus on the text aspect of code (e.g. programming problems \cite{apps, humaneval}). Recent work has shown that screenshots can help self-revise front-end code \cite{design2code}. Our work further extends program repair into the visual domain by showing how we can isolate visual differences in a layer-wise way that is specific to the layer-based organization of design elements on a canvas. Additionally, we show how visually-grounded program repair can be both automatically and interactively integrated within a user interface.

\subsection{Animation Tools for Novices}

Many prior works have introduced approaches for creating animation from static assets. Often, these approaches scope around specific artifacts such as character animation, dynamic illustrations, explainer videos, animated unit visualization, and 3D animation \cite{Dontcheva:03:Layered, Kazi:2016:MotionAmplifiers, Jahanlou:22:Katika, Cao:23:DataParticles, Ma:Stylized3D}. Systems focused on animated storytelling have explored how stories and scripts can be converted into animation. DataParticles \cite{Cao:23:DataParticles} takes in stories and outputs animated unit visualizations, animating chunks of story with animation effects that connect text and visuals. Katika \cite{Jahanlou:22:Katika} takes in scripts and outputs amateur explainer videos, animating shots with motion bundles and SVG assets. While these systems show how animation can be guided by natural language, the mappings they make between language and animation are preset-based (e.g. ``pop and pulsing", ``highlight") and can be too limited in expressive range for logo animation.

Other HCI systems have explored how other modalities of input like sketch, texture, performance, deformation, and video \cite{KSketch, Dontcheva:03:Layered, Kazi:Draco, Igarashi:05:AsRigid, Joshi:12:Cliplets} can drive animation. Draco and Kitty show how sketch input and kinetic textures can create path and particle motion for digital illustration \cite{Kazi:KineticTextures}. Motion Amplifiers and Ma et. al. show how presets implementing animation design principles (e.g. squash-and-stretch, follow-through) \cite{Disney:81:PrinciplesAnimation} can help users create stylized 2D and 3D animation \cite{Kazi:2016:MotionAmplifiers, Ma:Stylized3D}. Performance-based systems explored how gestures and acting can create interactive graphic overlays and layered character animation \cite{Dontcheva:03:Layered,Saquib:19:Interactive}. Recent work such as DrawTalking \cite{Rosenberg:24:DrawTalking} show how different modalities like sketch and speech can be combined during animation. Mixed-initiative interfaces for animation \cite{Willett:18:MixedInitiative, Gunturu:24:AugmentedPhysics} show that users can benefit from automatic support for tedious tasks such as object segmentation. However, the freeform nature of inputs like sketch, texture, or performance can be underconstrained for a professional use case like logo animation, and many of these prior approaches still primarily work around parameter-based editing. Lastly, other approaches allow users to transfer motion from input videos or source animations \cite{Zhang:2023:Motion, Joshi:12:Cliplets}. However, these approaches make users start from a separate source of content and work backwards, which can be ill-posed for something as content-specific as logo animation.

\subsection{Generative Tools for Animation}

Generative models have introduced new approaches for animation. Techniques such as frame interpolation and embedding-based interpolation have provided new techniques and tools for creating stylized animation and morphing \cite{Niklaus:2017:FrameInterpolation, discodiffusion,Chen:2023:Seine}. Models like AnimateDiff and motion LoRAs allow users to generate animated videos using prompts and learned motion effects \cite{Guo:2023:Animatediff}. Tools such as Runway Motion Brush \cite{Runway:MotionBrush} allow users to brush over parts of an image or video to animate. While these aforementioned tools are powerful at creating motion that matches the underlying content, they animate in pixel space and treat image and video as one flat layer, which make them ineffective at handling layered inputs such as layouts. \nickname shows how AI tools can handle inputs with layers by automatically grouping them and letting users interactively edit this organization through layer panel interactions.

A separate line of work looks at generating code for motion design and animation. Spellburst \cite{Angert:23:Spellburst} was an early work showing that users can generate creative code (p5.js) and edit by merging code representations. Keyframer \cite{Tseng:24:Keyframer} was another work that explored how novice and expert designers can generate CSS animation with LLMs and iterate through prompts and direct editing of code. This work also described that users had difficulty controlling the grouping and timing of animation using natural language. Both of these generative approaches operated without visual context--they understood the generated animation only in terms of its text representation (e.g. Keyframer used short text labels to describe the elements being animated). \nickname builds on these two works but is the first to leverage VLMs for animation and use visual analysis to inform code implementation. \nickname also introduces code-connected AI editing widgets, which let users edit animation at a higher level of abstraction than code and with more targeted control and intuitive GUI affordances than prior approaches \cite{Tseng:24:Keyframer, Angert:23:Spellburst}.
% , to provide more specification and control support during editing.
% . at the high-level (through prompts and high-level steering quick actions), with local specificity (prompts bound to animation events), and the low-level (direct interaction with a timeline). These provide more efficient and effective interactions for natural language authoring of animation than prompts alone. 
% We compare our approach against a state-of-the-art industry tool across a wide range of use cases and show a significant improvement in content awareness. 

% Works such as DirectGPT \cite{Masson:24:DirectGPT} point out that there is a need to find abstractions that help prompts interface better with GUI for more efficient interaction. 

% These ideas narrative timeline is also a practical improvement upon traditional timelines, which are generic blocks that do not describe the events within them. The narrative timeline can generalize to other animation / video editing tasks. 
\section{Formative Steps}

\revAdd{To understand the scope of logo animation, we took a mixed methods approach: 1) interviewing motion designers and 2) analyzing existing end-user tools. We first conducted interviews with experts to inform our approach. Then, we additionally analyzed end user tools to understand the design patterns commercial tools offer novices to animation.}

\subsection{Formative Methodology}

\revAdd{We interviewed motion design professionals (E1-E4), each of whom had at least 10 years of experience in logo and brand animation. E1, E2, E4 were recruited from a freelancing platform; E3 was recruited from within a design company. Each designer was interviewed and compensated for an hour of their time, during which they presented their 2-3 projects from their portfolio and spoke to their design process.}

\subsection{Formative Insights around Motion Design and Logo Animation}

\subsubsection{\revAdd{Motion designers strive to make the animation semantically meaningful.}}

\revAdd{Motion designers often begin the animation process by receiving concepts, design briefs, and brand books from clients. These documents often help tie the important branding elements of the logo with semantic context about brand identity. For example, E1 described how clients would ask for them to make the text within the logo animation move playfully or have the graphic elements rain down the screen with Tetris-like game mechanics to create a nostalgic feel. }

\revAdd{E4 described how it was especially important that the motion on the primary element match how that element might visually be expected to move. They stated, \textit{``For logos, it is the brand identity. If it's a tree it needs to grow. If it's a wave, it needs to be waving. It has to be something specific to the logo."} They described animated templates as a design solution they know clients would try but added context about the limitations to templates: \textit{``It's hard to generalize it [refers to a logo animation template]. You can just swap the logo in and out, but it has nothing specific to it.}}

\subsubsection{\revAdd{The design of logo animation is influenced by visual qualities such as layout, symmetry, layer relationships.}}

% Motion designers emphasized that the animation also has to work with the visual nature of the design elements and their arrangement on the canvas.

\revAdd{E2 and E4 both showed animated logo reveals from their portfolios that had a \textit{``zero-to-hero"} effect. In these, the primary element, the logo, often gets a hero moment, while the rest of the secondary elements would enter this hero moment with supporting animation. In this way, the \textit{visual hierarchy} of a layout is a source of information that can inform the amount of motion applied on each element. Other sources of information include \textit{positioning} and \textit{grouping}. E1 showed a project where they animated a dotted border by staggering a fade-in effect across all the dots. Creating groups and applying animations to groups creates a sense of coordination and visual flow. E2 also showed a project where they animated two halves of an infinity logo to move symmetrically into place, showing how animation can reflect visual relationships like symmetry. }

% On top of these aspects of layout, we also found through experimentation that because logos are often heavily stylized, the motion can also be stylized to the point where the most important quality to get right property can be the direction of the motion. Whether an object moves in off-screen from the left, right, top, bottom, or from the origin derives from the \textit{orientation} of the graphic element--whether it appears flat and front-facing or oriented towards one direction.

\subsubsection{Professionals focus on visual flow and the technical details of motion dynamics.}

Professionals like E2 stated that they spent most of their time getting the visual flow right. This meant experimenting with different timings and easing functions. It can be hard even for an expert to get these aspects of motion dynamics to feel right, suggesting that this could be a tedious aspect of animation that would benefit from automatic support. \revAdd{E2 showed timelines they worked over in After Effects, which had dozens of keyframes that they kept track of by repeatedly playing back the animation as they created it.}

\subsection{Analysis of End User Tools for Layout Animation}
\revAdd{To complement our expert interviews, we analyzed commercial tools novices use for animation. These tools included Adobe Express, Canva, Google Slides, Capcut, and Pinterest Shuffles \cite{AdobeExpress:AnimatedTemplates, Canva:MagicAnimate, AdobePremiereTemplates, Capcut, PinterestShuffles}. In these tools, there are often animation panels with motion presets, design galleries of animated templates, and automatic animation techniques. Motion presets are familiar effects like fade, wipe, and dissolve that novices can edit in terms of duration, speed, and direction. In tools like Adobe Express, Canva, and Google Slides, a direction parameter often sets simple path motion like an element sliding in from left on entrance or exit. Some tools such as Powerpoint and Canvas allow users to define path motion with sketch input or preset paths. In Adobe Express, Canva, and Powerpoint, tools generally provide users with 1-2 dozen presets, often organized by motion labels like ``energetic" or ``professional". These tools often restrict novices to working within a few states such as on entrance, on main duration, looping, or exit \cite{intro_looping_outro}. Tools also offer galleries of animated templates, which have prebaked motion and placeholders users can swap out with their own content \cite{AdobeExpress:AnimatedTemplates}. Lastly, tools such as Figma and Canva offer automatic animation techniques, so that novices do not have to individually animate each element. Figma Smart Animate and Powerpoint Morph find matching layers across slides and use property differences to automatically set animation \cite{Figma:SmartAnimate, Powerpoint:Morph}. Canva Magic Animate animates layouts based on an input motion style and a computational understanding of the canvas layers \cite{Canva:MagicAnimate}. }

From these formative steps, we derived the following guiding design principles:

\revAdd{DG1) \textit{To create semantically meaningful animation, an AI approach should generate code that users can easily edit and engage with in natural language.} Users can benefit from having a design concept that guides the code implementation, which they can edit later on.  }

% make it map to visual grounding
\revAdd{DG2) \textit{To create animation customized for the design, an AI approach should apply visual analysis to guide the code implementation.} Users should be able to edit how an AI organizes the animation in terms of which elements get more emphasis and how elements group within the animation.}

% Automatic analysis of what elements should have more visual precedence and what layers can group for coordination can produce better animation. 

\revAdd{DG3) \textit{To help novices edit their animation, an AI approach for animation should help visualize the animation and in what ways it can be edited.} Users should be able to understand the events that make up their animation and easily identify aspects that they can edit such as timing, emphasis, and speed.}

\revAdd{DG4) \textit{To prevent novices from having  to understand the technical details of the code, an AI approach for animation should help users handle the implementation details.} Users should be able to edit with high-level controls in the form of GUI support that allows them to make targeted edits to their animation.}
\section{\nickname System}
% \rubaiat{In this section, we've often used specific assumptions and constraints (eg, grouping categories, length of animation) specific to logo animation. Should we call this out explicitly? Readers might be curious to know how the overall approach can be extended to other types of artifacts (eg, layout), that have different grouping, time constraints, and motion synchorization rules. Finally, we haven't talked much about high-level controls / parameters.}

\begin{figure*}
\centering
\includegraphics[width=\textwidth]{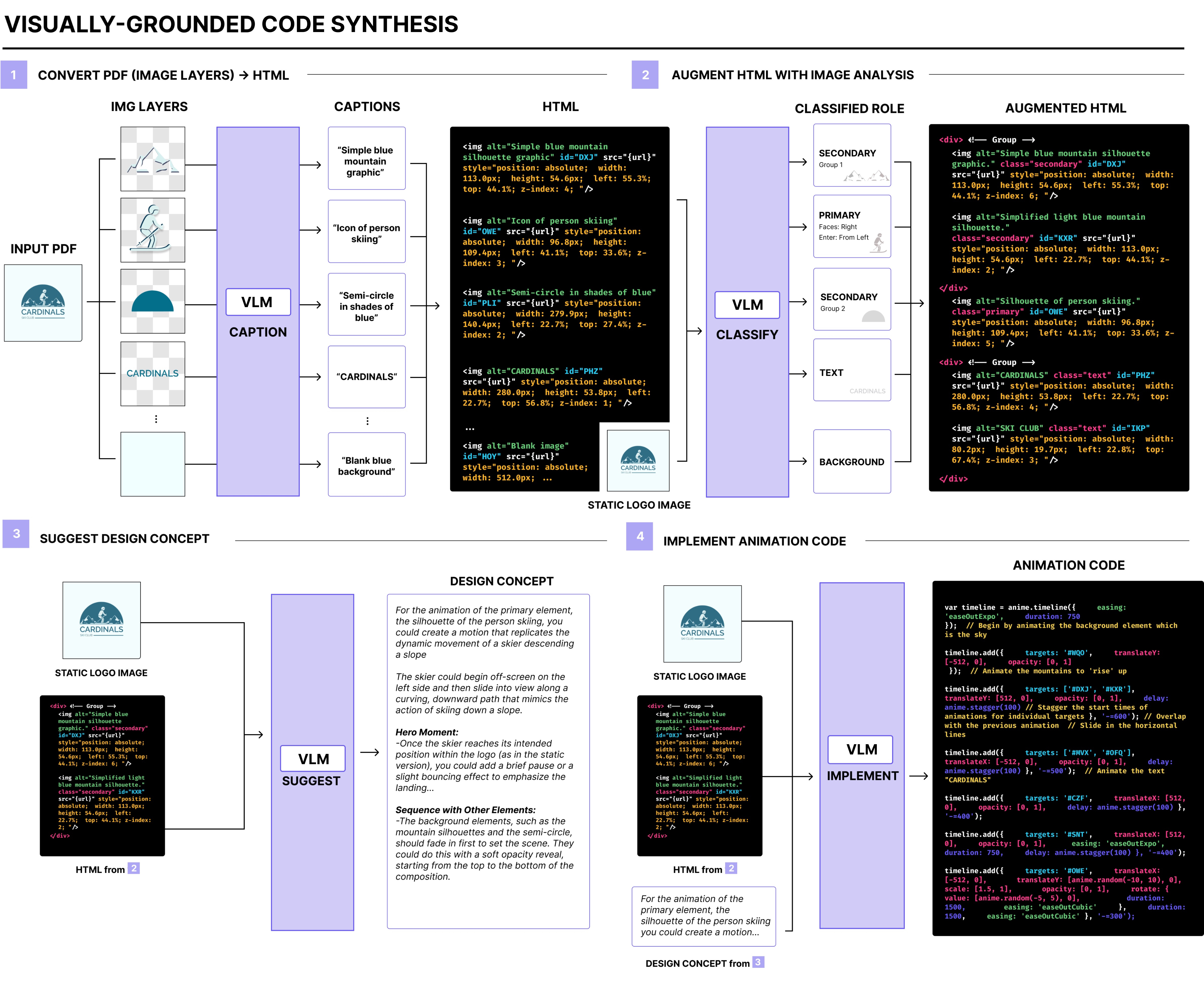}
\caption{Visually-Grounded Code Synthesis Overview. In Step 1, a PDF of a logo is converted into an HTML representation of the canvas. In Step 2, \nickname augments the HTML with information from visual analysis steps. In Step 3, \nickname suggests a design concept for the animated logo. In Step 4, a VLM implements animation code for the design concept. This code animates the logo HTML. }
\label{fig:synthesis_overview}
\Description{Diagram showing four image layers on the left being assigned a caption by GPT4-V. 1) CONVERT PDF IMAGE LAYERS TO HTML. For example, an icon of a skiier is mapped to the caption "silhouette of person skiing". Next to these captions there is an HTML snippet and an image preview, which shows the new pair of multimodal inputs to the next stage of the LLM. 2) AUGMENT HTML WITH IMAGE ANALYSIS. The conceptual groups show that secondary elements are captured within divs and all roles (primary, secondary, text, and background are tagged. 3) SUGGEST DESIGN CONCEPT shows how the static logo and augmented html from Step 2 are used to suggest a design concept. 4) IMPLEMENT ANIMATION CODE uses the design concept, image, and HTML to implement animation code.}
\end{figure*}

\revAdd{We present \nickname, an AI code generation and editing tool that helps novices create semantically meaningful animation. We walkthrough \nickname's approach with a Ski Club logo as a grounding example. First, \nickname helps novices automatically animate the logo by generating code using \textbf{visually-grounded code synthesis and program repair}. This method takes in a logo in the form of a layered PDF document as input and generates an HTML page and animation code as output. The animation that executes on the page is semantically meaningful to the logo design. For example, a skier element can ski in from the left and do a flip, and the title text can pulse in just as the skier lands. Second, novices can customize this animation using \textbf{code-connected AI editing widgets} we introduce such as a narrative timeline, layer panel, and quick actions. These code-connected AI editing widgets implement edits to the motion, grouping, and timing of the animation by doing targeted regeneration of the underlying code. For example, a user can edit the skier to have a different hero moment (e.g. skier skis in diagonally as if on a slope) or synchronize the timing of all secondary elements with the skier's entrance. We now walk through how users can create and edit animation with \nickname. }
% Using these familiar GUI controls, novices can make targeted edits to the motion, grouping, and timing of their animation. 
% As output, LogoMotion produces an HTML page with JavaScript code that runs the animation on the logo. Users can then edit this animation using code-connected AI widgets. 
% As output, \nickname automatically generates an HTML page with Javascript code that produces animation on the page. 

% automatically finds and self-debugs code errors in the animation by comparing how the animation ends with the original layout. 

% but we discuss the generalizability of our approach to other layered documents (e.g. social media posts, banners) at the end of this section.

\subsection{Input}
\revAdd{As input, \nickname takes in a PDF that can consist of multiple image and text layers. Fig. \ref{fig:synthesis_overview} shows an example. The logo has a skier, two mountains, decorative lines, a colored background for image elements, and ``CARDINALS" and "Ski Club" for text elements. This PDF is a layered design that has already been arranged by a designer, so each element has a pre-defined position and a layer index. \nickname starts with all layers not attached to any groups.}

% \revAdd{\nickname's input is a PDF representing the static design. \nickname can take in any kind of static design but we focus on logo layouts (e.g. Fig. \ref{fig:synthesis_overview}), which generally have one primary graphic element, one text element, and secondary graphic elements.}

\subsubsection{Preprocessing} 
\revAdd{
To create a code representation of the canvas, \nickname uses a custom-written Extendscript program that takes a PDF imported in Adobe Illustrator and exports it into an HTML page. The Extendscript iterates through the text and image layers and converts each one into an HTML element with a unique ID (``\#EHP"). It copies over bounding box information and z-order indices into the style attribute of each element. Fig. \ref{fig:synthesis_overview} shows how the skier image and text elements are exported as HTML. For text elements, \nickname also copies over the text content in alt text (alt=``SKI CLUB") and sets a style attribute (class=``text"). This HTML representation helps \nickname understand how to target image and text layers in the animation code.}

\subsection{Visually-Grounded Code Synthesis}
\label{sec:method:preprocess}
\revAdd{During visually-grounded code synthesis, \nickname performs visual analysis to understand the semantics of the logo and instantiates a design concept to inform the implementation of animation code.}

\subsubsection{Layer Captioning}
To understand what each element represents, \nickname first captions each layer. Fig. \ref{fig:synthesis_overview}-Step 1 shows how a skier image is captioned as an "icon of person skiing". To caption, \nickname isolates each layer against a plain background, queries a VLM for a caption, and sets the caption to be the alt text of the  HTML element. 

% This background is white by default or neon green for low-contrast elements. 
\subsubsection{Visual Hierarchy} 
\revAdd{\nickname analyzes the logo image to understand the visual hierarchy of the elements within it. This visual analysis can define which elements should get more emphasis (the hero moment in a logo) and which elements should get less (secondary supporting motion). \nickname passes the HTML code (which holds information about the element's caption, sizing, positioning, layering) and the logo image through a VLM call that outputs a JSON classifying each HTML element as primary, secondary, text, or background. These role classifications are added to each HTML element as a class attribute (e.g. class=``primary''). \nickname chooses only one element as the primary element and applies additional visual analysis to understand which way it faces and how the element should enter. For example, if the skier element appears to face the right side of the canvas, it should ski and flip in from offscreen left (Fig. \ref{fig:synthesis_overview}-Step 2).}

\subsubsection{Grouping Elements}

\revAdd{In an animation, elements that conceptually relate ideally move in a coordinated way. To support this, \nickname automatically makes groups over the secondary elements and applies animations to groups. It does this by taking in the HTML and logo image and outputting a JSON that places layers in groups (e.g. secondary-group-1: \#EHY, \#OWE). Elements in the same group are then reorganized in the HTML to be under the same parent div element (e.g. <div class="secondary group-1">). In Fig. \ref{fig:synthesis_overview}-Step 2, we show how \nickname groups the two mountain elements, two line elements, and two text elements together. There can be any number of groups and grouping is especially effective a layout has many secondary elements, such as a border of repeating circles or a dozen stars. }

\subsubsection{Design Concept}

% describing how each group of elements animates and in what sequence.
With an HTML representation of the logo that is augmented with visual analysis, we are ready to animate. To create a semantically meaningful animation, \nickname instantiates an animation \textit{design concept}. The design concept is a pseudocode that semantically conditions the code generation. It suggests how each group of elements should animate, what hero moment can be applied to the primary element, and how all the elements sequence and time in. It can be multiple paragraphs, as shown in Fig. \ref{fig:synthesis_overview}-Step 3: \textit{``The skier could begin off-screen and slide into view along a curving, downward path that mimics the action of skiing down a slope. Once the skier reaches its intended position, add a brief pause or slight bouncing effect to emphasize the landing."} To generate the design concept, we provided a VLM with 1) the HTML from the previous step and 2) logo image. This is an example of our prompt:

\begin{lstlisting}
This image is of a logo that we would like to animate.
Here is the HTML representation of this logo: <HTML>

We want to implement a logo animation which has a hero moment on the primary element. The primary element in this caption should animate in a way that mimics its typical behavior or actions in the real world. 

We analyzed the image to decide if it in its entrance it should take a path onto the screen or not: <entrance description>. Considering this information, suggest a motion that characterizes how this element ({primary element image caption}) could move while onscreen.

Additionally, suggest how this element should be sequenced in the context of a logo reveal with the other secondary, text, and background elements. (Note that the element is an image layer, so parts within it cannot be animated.)
\end{lstlisting}

% \begin{itemize}
% \list The primary element should have a engaging animation that is relevant to the subject matter of the logo and what it depicts.
% \list Secondary elements can get simpler animation, to not detract from the primary element's animation.
% \list Elements that are underneath other elements come in earlier.
% \list The background element should NOT be animated.

\subsubsection{Code Implementation} 
\revAdd{\nickname then implements the animation code using the 1) HTML that has been informed by visual analysis, 2) design concept, and 3) logo image. The prompt instructs the VLM (GPT-4o) to initialize a variable for a timeline and write animation events that target groups (HTML classes) and elements (HTML IDs). The generated code is then run on the HTML page, which includes anime.js as a library and presents all the layers on a canvas. \nickname can also be adapted to write in other animation frameworks (e.g. CSS animation). An example of how \nickname translates a description from the design concept into animation code is provided below.} 

\begin{lstlisting}
timeline.add({
    targets: '#OWE', \\ Skier enters from the left, does a spin trick in midair
    translateX: [-512, 0], \\ Moves across from the left
    translateY: [0, -100, 0], \\ Jumping effect
    rotate: ['0deg', '360deg'] \\ Spin trick
    easing: 'easeInOutSine', \\ Easing for smooth animation
    duration: 1000, startTime: 3000}
\end{lstlisting}

\subsection{Visually-Grounded Program Repair}

\begin{figure*}
\centering
\includegraphics[width=\textwidth]{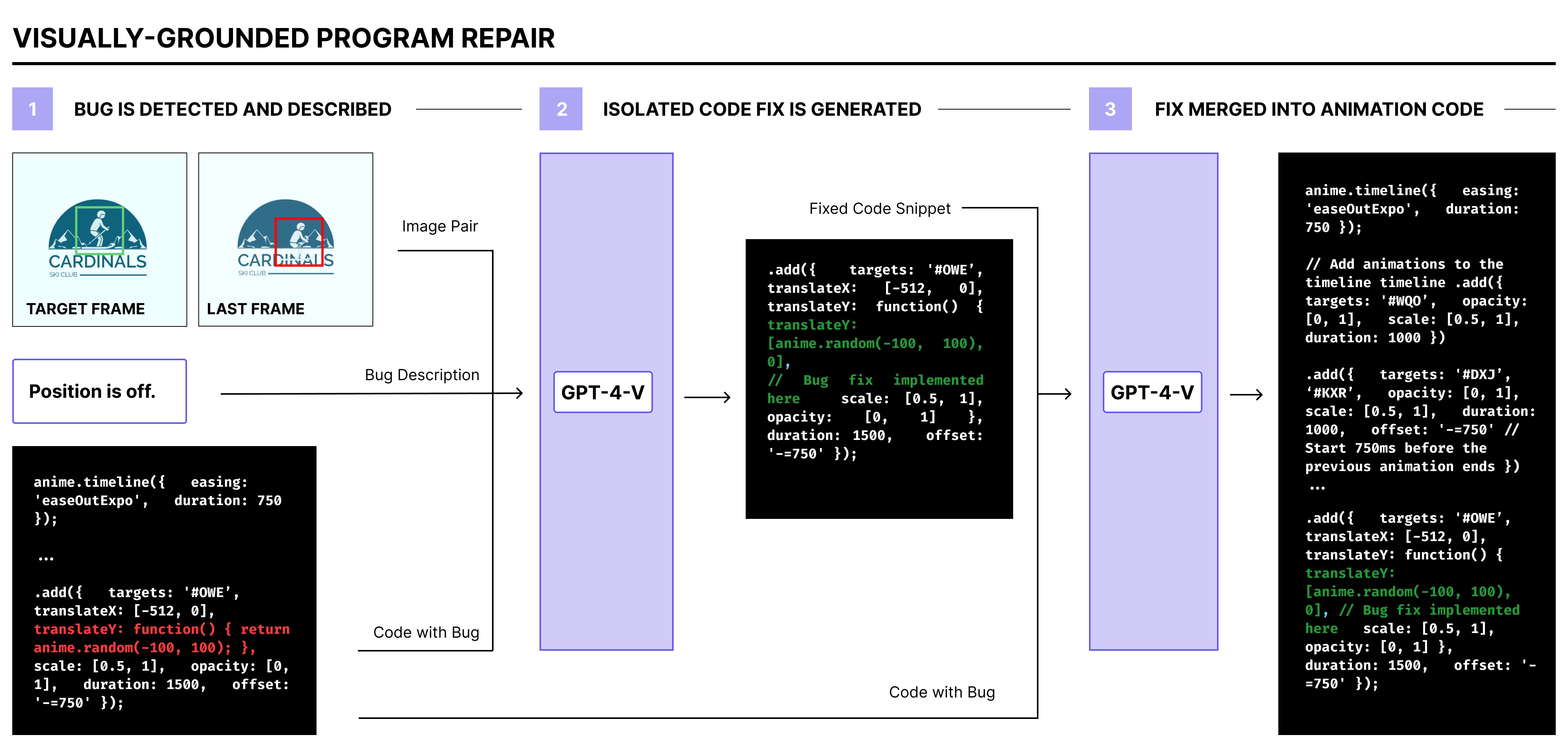}
\caption{Visually-grounded program repair takes visual feedback from the canvas to self-debug animation code errors. It identifies bugs in animation code by checking for differences between elements in the target layout and elements in the last frame of the animation. If there is a visual error, a VLM receives 1) an image pair of the element with an error (the rest of the layout behind the skier element is pictured only for context), 2) a bug description, and 3) the original code. It outputs a code fix that is merged back into the animation code.}
\label{fig:visual_debugging}

\Description{Visually grounded program repair. 1) BUG IS DETECTED AND DESCRIBED. The first step takes in two images as input, a description of the bug, and the code that triggered the error. The diagram has a pair of images at the top, one of which annotates a position error with red and green bounding boxes. The image pair, but description, and code bugs are inputs to the next step, where GPT-4-V generates 2) an isolated code fix. This fixed code snippet is merged back into the animation code in the third step. }
\end{figure*}

\revAdd{AI-generated code can have errors and animation code in particular can have errors that render visually. For example, generated animation code can leave the skier element twenty pixels off from the intended position, because of an incorrect assumption AI makes about array syntax. To catch and fix these errors, \nickname use visual feedback from the canvas for self-debugging in a method we call \textbf{visually-grounded program repair}. }

\revAdd{Once the animation code from the previous code synthesis stage is generated, it is automatically executed on the HTML logo. After the code is run, \nickname checks the last frame of the animation. It steps through each layer and programmatically compares the position, scale, rotation, and opacity for what it should be in the original logo layout. It tests for equivalence on attributes from the bounding box coordinates  (left, top, width, height) and style properties (opacity). Because VLMs often make mistakes when provided unnecessary visual context \cite{Bang:2023:Multitask, Favero:2024:Multimodal}, \nickname goes layer by layer to reduce the information that the VLM needs to process. }

\revAdd{Once an error is detected, we pass in the element ID(s), a bug description, animation code, and a pair of images capturing the difference. This image pair shows the element on its own--the rest of the layers are hidden before \nickname takes a screenshot. The images are labeled by ``TARGET FRAME" and ``LAST FRAME". Fig. \ref{fig:visual_debugging} shows an example of a bug that renders visually: the skier is off position. \nickname then generates a code snippet to fix it and re-executes the animation with the proposed fix. If the animation error is fixed, the snippet is merged back in with a VLM call. We allow \nickname to try fixing up to four times. Though this step catches important errors that violate the original layout of the logo, it does not check for all possible errors (e.g. layer occlusions in intermediate frames). With the conclusion of checking, \nickname completes the automatic stages that generate a logo animation.}

\subsection{\revAdd{Editing with Code-Connected AI Editing Widgets}}

\begin{figure*}
\centering
\includegraphics[width=\textwidth]{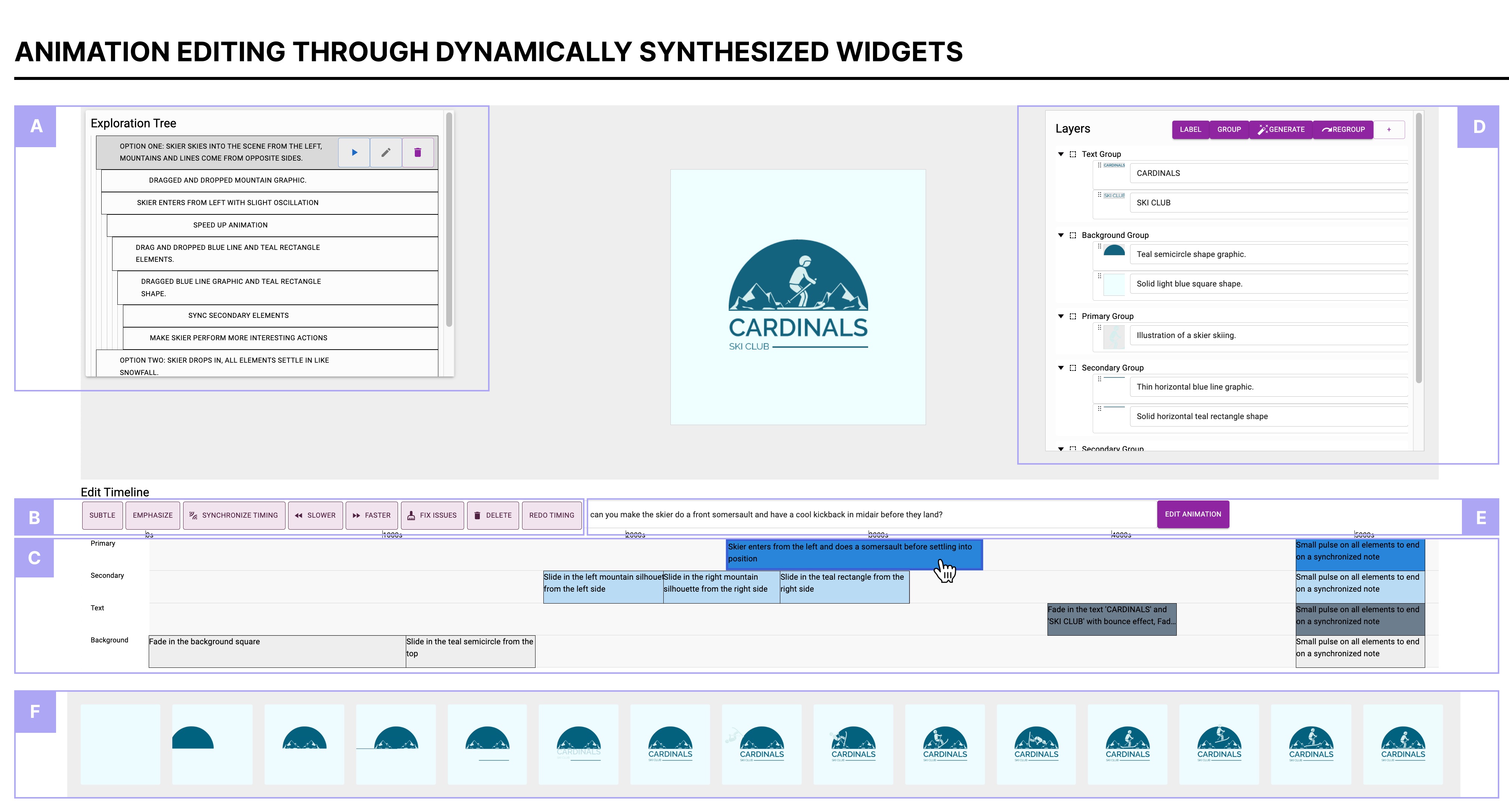}
\caption{LogoMotion provides users code-connected widgets to edit their animation. It provides quick actions (B), a narrative timeline (C), a layer panel (D), and prompt-based interactions (E). These widgets help users make targeted edits to the motion, grouping, and timing of their animation. An example animation edit output is shown in (F). }
\label{fig:interface}
\Description{LogoMotion's animation editor and code-connected widgets are depicted. The Exploration tree is in the top left, where it shows a version history of animation edits. There is a canvas in the center showing the skier logo. There is a layer panel with labels and groupings assigned by LogoMotion in the top right. There is a row of pink buttons which refer to the quick actions. There is a prompt box where the prompt says, ``can you make the skier do a front somersault and have a cool kickback in midair before they land?" Below a narrative timeline shows how the animation to be edited is being modeled by LogoMotion. It has four tracks with a different color for each track. Below that a row shows an animation that implements that user edit request.}
\end{figure*}

\revAdd{\nickname allows novice designers to edit their animation using \textbf{code-connected AI editing widgets} within the UI pictured in Fig. \ref{fig:interface}. The UI includes a canvas for viewing the animation, an Exploration Tree that tracks history, and the following code-connected AI editing widgets: 1) a narrative timeline, 2) a layer panel, 3) and a quick action panel. We call these widgets \textit{code-connected} because when users make edits with them, \nickname implements the edit by regenerating part or all of the underlying animation code. We enable this connection between the editing widgets and the AI-generated animation code by processing the HTML code representation into the groups of the layer panel (Fig. \ref{fig:interface}-D) and the Javascript animation code representation into the narrative timeline (Fig. \ref{fig:interface}-C). The intuitive affordances of a GUI representation helps users edit the motion, grouping, and timing of the animation at a higher level of abstraction over the underlying code.} 

\subsubsection{Narrative Timeline.}

\revAdd{The most prominent widget is the narrative timeline. It ``narrates" what elements are doing during the animation. The timeline visualizes four tracks for Primary, Secondary, Text, and Background that populate with animation blocks. Each block is labeled with a description of the animation events that take place within the span of the block. For example, a block mapping to a group of mountains on the Secondary Track can have the label \textit{``mountains rise in"}. The narrative timeline allows users to read the design concept that was generated in natural language rather than having to understand code. These labels can also be edited by users to change the animation events within that block. To create the narrative timeline, \nickname extracts the code comments attached to each block of animation code and uses them as labels. \nickname also defines the width and position of the block on the timeline by finding the start and end timestamps of each element and doing a merge-and-sort over all elements in a group. Each animation block captures metadata about the code, such as what code block it maps to, what elements it targets, and what its timestamps are.  
}

\revAdd{Users can write prompts to edit the animation, and these prompts can be bound to selected animation blocks. For example, Fig. \ref{fig:interface}-E shows how a user can prompt \textit{``can you make the skier do a front somersault..."} and select the skier block as shown in Fig. \ref{fig:interface}-E. Clicking \textit{Edit Animation} triggers \nickname to regenerate the code, specifically targeting that block. Users can also select multiple blocks at once and write a prompt. For example, they can write \textit{``make them all fall in like snow"} and have the edit apply across multiple selected groups. Lastly, the narrative timeline supports drag-and-drop, so users can edit start and end timestamps. Drag-and-drop requests are also routed to the VLM to implement, as exact start and end timestamps can be dependent upon other parameters like offsets (``+=200ms") and are not as simple to replace programmatically.  For every prompt edit, \nickname uses GPT-4o and takes in the following inputs: block metadata, the animation code, HTML, and the logo image.}

\subsubsection{Layer Panel to Reorganize Animation.}
\revAdd{\nickname automatically decides groups during the code synthesis, but users should be able to edit these groups. \nickname lets users do so with a layer panel (Fig. \ref{fig:interface}-D) that reflects the groups of the underlying HTML representation. Users could rename layers to have new identifiers, rearrange layers to be in new groups, create a new group, or delete groups. Clicking \textit{Regroup} would trigger \nickname to update the HTML code representation by calling a VLM to regenerate the animation code and reorganize it around the new groups. For example, at first, a user could have mountains, lines, and the semicircle group all doing different animations. The user could put them all within the same group and request a \textit{Regroup} which would regenerate the code to make all elements in the same group synchronized to have the same timing and motion.}

% LogoMotion provides users with quick actions to edit the synchronization, speed, and amount of animation  on elements. Quick actions included:
\subsubsection{Quick Actions}
\revAdd{To provide users with high-level controls and help users identify ways they can edit the animation, \nickname provides quick actions: \textit{Subtle, Emphasize, Synchronize Timing, Slower, Faster} and \textit{Delete} (Fig. \ref{fig:interface}-B). \textit{Slower / Faster} helps users test different speeds, and \textit{Subtle / Emphasize} helps users try different amounts of emphasis. \textit{Synchronize Timing} allows users to synchronize animations to align with the same timing. \textit{Delete} helps users identify and delete events within the animation. We derived this set of operations through experimentation and pilots with users, finding these to be the commonly requested actions. Quick actions can apply to selected blocks or to the entire animation timeline if no blocks are selected.  }

\subsubsection{Interactive Version of Program Repair.}

\revAdd{Every time the code is regenerated, \nickname automatically runs program repair. In pilots, we found that users sometimes did not want to wait for program repair to complete, because it could take a few attempts. They wanted to be able to stop it when it was unnecessary, so we allowed them to exit out of the loop. We also limited the checking to the primary element. As an alternative, we provided a \textit{Fix Issues} quick action to allow users to trigger visually-grounded program repair whenever they wanted. }

\subsubsection{Exploration Tree.}

\revAdd{To help users track version history, \nickname provides an Exploration Tree (Fig. \ref{fig:interface}-A). Users can choose one of four of \nickname's automatic animations and start editing. Each edit made becomes a new node on the Exploration Tree, and each node is labeled with what the edit was about (e.g. \textit{``add oscillation to skier entrance"}). Each time a user clicks on a node, \nickname updates the narrative timeline with the animation code and the layer panel with the layer and groups at that version.}

\subsubsection{Code-Connected Widget Implementation Details}

The system was implemented in Python / Flask and React/ Javascript. For VLMs, \nickname calls GPT-4v (for Evaluation 1 and 2) and GPT-4o (for Evaluation 3). \nickname stores the history of animation runs and repair history behind each run in JSON files. To implement visually-grounded program repair, \nickname copies each element over to a duplicate canvas that is hidden from the user, so that it can programmatically check layers and screenshot them without the user seeing any visual side effects on their canvas. \nickname checks the style attributes on each layer with a 1 percent tolerance level on pixel differences, so that program repair does not overtrigger on minute differences.

\bigskip
Through this approach, \nickname allows users to automatically generate semantically meaningful animation using \textbf{visually-grounded code-synthesis and program repair} and interactively customize the animation using \textbf{code-connected AI editing widgets}.

% Each animation block has metadata attached that maps it to the corresponding snippet of code and element IDs it targeted. This metadata was passed to the VLM whenever an edit request was made.

% To keep track of state, each animation and its program repair revisions were written into a JSON file. This file was processed to only play the most successful run (the run which debugged all elements, or the original run, if program repair did not resolve issues).

% \subsubsection{Narrative Timeline Derivation.} While the generated code could include information about start times and durations, the offsets and delay values made calculating a timeline from the code representation misleading. Instead, to capture the exact timings, we injected a callback into each animation event that would fire a) when it began and b) when it concluded. Additionally, each animation event included parameters like targets, duration, and offset which we parsed to create the final metadata which rendering each timeline block. From these element-level timings, we merged and sorted to aggregate group-level timings.

\color{black}
\section{Comparison Study on Automatic Outputs}

% We conduct two evaluations with novices and experts to establish the usability and usefulness of our system. To evaluate the effectiveness and robustness of our approach on a diverse set of templates, we recruited novices to try \nickname, judge the generated animations in terms of content awareness and system in terms of customizability).
We conducted three evaluations to understand the quality of \nickname: 1) a comparison study against an industry standard informed by professional logo animators 2) an empirical analysis of program repair testing different experimental settings, 3) an evaluation with novices (n=16) to understand \nickname's support for animation editing. These evaluations centered around the following research questions:

\begin{description}
    \item[RQ1:] To what extent does \nickname generate animation that is relevant to design elements on a canvas?
    % \item[H2:] Compared to the industry baseline, after a few rounds of interactive feedback, LLM-generated templates would be on par in terms of \underline{sequencing} on average.
   % \item[RQ2:] In what ways can \nickname support the customizability of generated animations through natural language interaction?
   \item[RQ2:] What are the overall strengths and the weaknesses of \nickname at animation?
   \item [RQ3:] What sorts of errors does \nickname tend to make in code?
    \item [RQ4:] How capably can visually-grounded program repair debug errors and what settings of program repair impact performance?
   \item[RQ5:] \revAdd{To what extent can \nickname's code-connected AI editing widgets help novices with animation editing? }

\end{description}

The first evaluation is a comparison study comparing \nickname with Canva Magic Animate \cite{Canva:MagicAnimate}. Magic Animate, is an AI-based tool for automatic animation that is one of Canva Pro's premium features. Magic Animate analyzes the design on a canvas and applies motion to match the content, font choices, images, and color. It provides a range of motion styles (``bold", ``professional", ``elegant") which can change the dynamics of the animation and cover a wide variety of use cases. We additionally created an ablated version of \nickname (\nickname-Ablated) to understand how higher-level VLM operators impacted the content-awareness of the approach. \revAdd{Specifically, we ablated Step 2 (visual analysis resulting in the Augmented HTML) and Step 3 (Design Concept) of our visually-grounded code synthesis stage. From experimentation, we found that some base amount of visual grounding (Step 1 - captions and layer information) was necessary for the approach to work, so we wanted to understand the effect of more versus less visual grounding.} Our hypotheses for RQ1 were the following.

\begin{description}
    \item[H1a:] Compared to the other conditions, \nickname-Full would produce animations that were more \underline{content-aware}.
    \item[H1b:] Compared to the other conditions, \nickname-Full would be improved in terms of \underline{sequencing}.
   \item[H1c:] Compared to the other conditions, \nickname-Full would be better in terms of \underline{execution quality}.

\end{description}

\subsubsection{Methodology}
We gathered a test set of 23 logo layouts that spanned different categories of objects (animate and inanimate), layouts, and use cases. Use cases included holiday greetings, school clubs, advertisements, and corporate branding. All logo layouts were sourced from Adobe Express and Canva and are accessible online. 

To understand the quality of the automatically generated animations, we ran the first two stages of \nickname to get four animations for each logo. We then ran \nickname-Ablated to get another four animations for each logo. Generating a set of four \nickname / \nickname-Ablated outputs took approximately 12 minutes. To gather four animations for Magic Animate, we took the motion style they recommended for the layout as one animation and had an external logo animator pick the three next-best motion styles. The combination of a professional animator's decision on top of the fact that Magic Animate is an industry tool (engineered to have a polished outcome and wide applicability to different kinds of content) ensured we had a strong baseline to compare against.

To evaluate the animations, three professional designers were recruited to rate 276 animations (23 templates x 12 animations per logo) spanning the three conditions. Designers were introduced to the task with a remote call, calibrated with good and bad examples to understand the rubric, and compensated for their time. Each animation was presented in randomized order and rated on a scale of 1-5 for each of the following dimensions: 1) Relevance, 2) Sequencing, and 3) Execution Quality. Relevance describes how relevant the animation is to the subject matter of the logo. It is a measure of how specific the animation is to the design elements of the logo. Sequencing was a measure of how well the animation was sequenced in terms of coordination and timing across elements. Execution Quality judged the animation for how well it was executed and if it had any flaws. \revAdd{Each designer rated all 276 animations and took on average 2.83 hours to complete the task.}

\subsubsection{H1a. Relevance }

We averaged across the ratings of three design professionals. \nickname was rated to have significantly more relevance to the subject matter of the animated logos than both Magic Animate and \nickname-Ablated (H1a, \nickname-Full: M = 3.05, $\sigma = 0.64$; \nickname-Ablated: M = 2.68, $\sigma = 0.58$; Magic Animate: M = 2.33 , $\sigma$ = 0.33, $p \leq 0.001$). From this we confirm H1a; \nickname-Full was the top condition in terms of content-aware animations.

\begin{figure}[h]
    \centering
    \begin{subfigure}[b]{0.38\textwidth}
        \centering
        \includegraphics[width=\columnwidth]{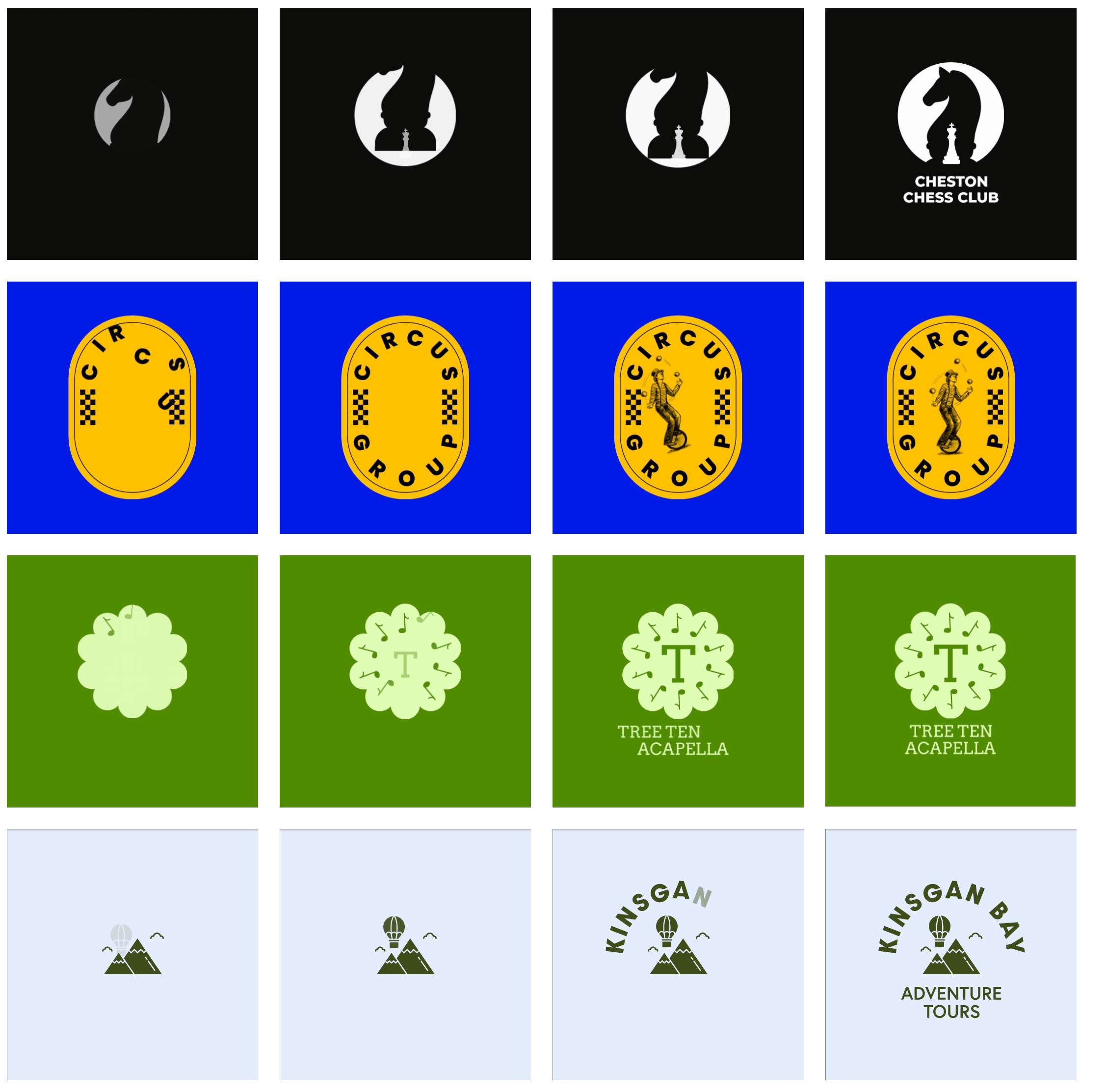}
        \caption{}
        \label{fig:logomotion_good}
        \Description{This 4x4 grid of images show frames from logo animations that were rated highly. The first row shows still frames for a chess animation, where text enters in the last frame. The second row shows elements for a circus animation coming in a staggered fashion, a unicycler comes in off-center and askew in frame three. In third frame, musical notes enter in a contralateral, symmetrical fashion as the text slide in from both sides. The last frame shows a hot air balloon that floats up behind a mountain as text comes in with a typewriter effect. The motion is described in the text. }
    \end{subfigure}
    \hfill
    \begin{subfigure}[b]{0.45\textwidth}
        \centering
    \includegraphics[width=\columnwidth]{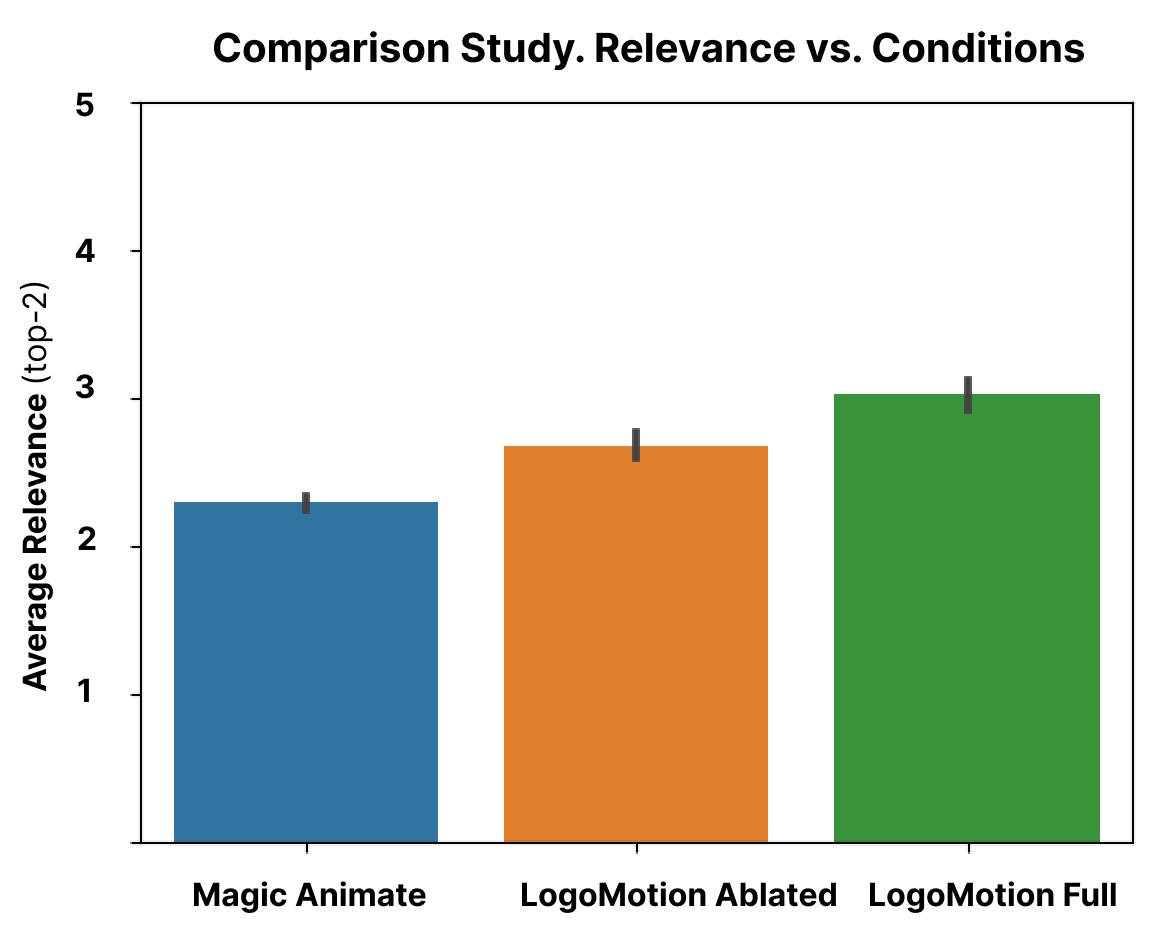}
    \caption{}
    \Description{Bar chart showing LogoMotion Full at the top, LogoMotion Ablated in the middle, and Magic Animate at the bottom. Values can be found in the Table 1. }

    \end{subfigure}

    \caption{\textbf{Comparison Study. a) } Examples of \nickname-generated logo animations. These animations show how \nickname is able to create motion that is characteristic of the design elements, layout-aware, and logically sequenced. \textbf{b)} We report the average top-2 relevance across conditions. \nickname was rated to be significantly better in terms of relevance to canvas content.}

    \label{fig:canva_comparison}
\end{figure}

When sorted by average relevance across raters, the top rated animations tended to come from \nickname (15 of top 20) or \nickname-Ablated (5 of top 20). \nickname animations that were rated highly showed semantically meaningful motion. The video for this paper shows examples of lanterns blowing as if in slight wind, crabs crawling zigzag into the screen, and a hockey stick swung as if in play. Frames from animations are depicted in Figure \ref{fig:logomotion_good}. In the first row, a black knight is translated into the canvas in an L-like motion as a bishop piece scales up. In the last row, we see a hot air balloon slowly rise in over the mountains, after which the logo title fades in letter by letter.

\subsubsection{H1b. Sequencing }

% Animations that were rated lower for Sequencing tended to have errors in layer ordering (an element closer to the top of the layers would come in before an element beneath it). However,

In terms of Sequencing, \nickname-Full was not significantly different from the other two conditions (H1b,  \nickname-Full: M = 3.15, $\sigma = 0.55$; \nickname-Ablated: M = 3.18, $\sigma$ = 0.41 ; Magic Animate: M = 3.12, $\sigma$ = 0.43). Qualitatively, \nickname-Full and \nickname-Ablated were both capable of implementing the logical sequencing of a logo reveal. They could time primary elements before secondary elements or vice versa, but generally put the text last. \nickname generally sequenced layers in from bottom to top. Animations rated lower for Sequencing tended to have errors in layer ordering. Background elements were often left static but could also be used effectively in animations. In the paper's video figure, we show that in an animated logo for a martial arts club, a silhouette of a karate character pops in from the left. As it lands into place, the background shakes, as if from the impact of a karate kick. The interaction between these design elements was suggested by the design concept: \textit{``for added impact, consider a screen shake or a vibration effect when the silhouette `lands' to draw even more attention to the primary element's hero moment."}

\nickname was also capable of generating animations that reflected properties of gestalt like symmetry. For example, the symmetrical elements of the Circus example in Fig. \ref{fig:logomotion_good}) animated in together, and the musical notes in the Acapella example animated in a contralateral, symmetrical way. \nickname could also create staggering within animation by mapping a generated animation function to a group of elements. This was how it implemented text animation such as a typewriter effect for arced text (Fig. \autoref{fig:logomotion_good}-row 4) and could stagger an opacity animation across a group of stars to create a twinkling effect.

% We did not measure the average length of an animation because that was dependent upon how many elements were on the canvas and how many conceptual groups were targeted in the animation code.

\subsubsection{H1c. Execution Quality }

\nickname-Full did not perform significantly differently from other conditions in terms of execution quality (H1b,  \nickname-Full: M = 3.25, $\sigma = 0.54$; \nickname-Ablated: M = 3.38, $\sigma$ = 0.46; Magic Animate: M = 3.22, $\sigma$ = 0.39). \nickname-Ablated scored the highest on execution quality. Many animations for \nickname-Ablated tended to be conceptually similar (all elements fade in or translate into place from a slight displacement). Animations for \nickname-Ablated tended to be minimal in complexity and thus easy to execute. What brought \nickname-Full down in terms of execution quality was that sometimes \nickname-Full animated box-shadows and background-color changes, which made it past our program repair checker and were left unrevised. Lastly, for context, Magic Animate could at times receive lower scores on execution quality, because some of the animations it assigned produced a cropping effect that was an undesirable visual side effect.

\subsubsection{Comparison Study Conclusion.} This comparison study illustrated that \nickname could outperform another AI-based industry standard animation tool at producing animation specific to the design elements on the canvas. \nickname's animations were comparable in terms of execution quality and logical sequencing.

% This interface allowed them to playback the animations and edit any previous run. Their  requests were implemented in code and merged with the relevant animation code timeline. They could also choose to edit from an intermediate version of the code (that they had prompted for, not from the original 4 runs).

\begin{table}
  \centering
  \begin{tabular}{lccc}
    & Relevance & Sequencing  & Execution Quality \\
    \hline
    LogoMotion Full  & \textbf{3.05}\textsuperscript{**} &   3.15 & 3.25   \\
        LogoMotion Ablated & 2.68  &  3.18  & 3.38\\
       Magic Animate &  2.33 & 3.12  & 3.22 \\
    \hline
  \end{tabular}
  \caption{Relevance, Sequencing, Execution Quality. Average expert designer ratings evaluating automatically animated logos (n=276) for three conditions. (\textsuperscript{**} $= p \leq 0.001$)}
  \label{tab:interrater-reliability}
  \Description{Table showing ratings for expert evaluation. LogoMotion was rated significantly higher in terms of relevance over Magic Animate, an industry standard tool for animation.}
\end{table}

\section{Technical Evaluation of Program Repair}
Next, we conducted an evaluation for visually-grounded program repair stage to empirically understand what errors \nickname tended to make (RQ3) and how capable it was at self-debugging them (RQ4).
% We provide empirical analysis of this stage testing different experimental settings.

% Specifically, in terms of research questions, we wanted to understand:

% \begin{itemize}
% \item RQ4. What sorts of errors does the program synthesis stage make?
% \item RQ5. How capably can visually-grounded program repair debug such errors and what settings of program repair impact performance?
% \item RQ5. How vary in performance across different types of errors
% \item RQ6. Does the performance of program repair vary depending on a) the type of error, b) the number of attempts allowed, c) number of elements in a design?
% \end{itemize}

\subsection{Methodology}

% \begin{figure}
% \includegraphics[width=0.25\textwidth]{fig/logomotion_errors.jpg}
% \caption{Examples of errors \nickname could make but also repair.}
% \label{fig:interface}
% \end{figure}

Sixty-eight percent of the animation runs from the outset (after the first stage of code synthesis) were error-free and did not require program repair. The other 32\% required the program repair. Within this stage, we modulated a 1) hyperparameter $k$ and 2) whether or not image context was provided. $k$ upper bounded the number of attempts a VLM could take to solve the bug and was varied from 1 to 4 attempts. Varying $k$ is modeled after the pass@k methodology proposed by HumanEval \cite{humaneval}, where $k$ code samples are generated in attempts to solve a problem and the fraction of problems solved is the solve rate. In this case, the pass@k framework is applied in the context of program repair.

The second setting that we varied was whether or not image context about the visual error was provided. This image context (pictured in Fig. \ref{fig:visual_debugging}) is a labeled image pair showing a layer at its target position vs. in its last frame in the animation. We refer to Repair$_{+Imgs}$ as the setting with both visual context and bounding box information about the error. We refer to Repair$_{-Imgs}$ as the setting where no visual context and only bounding box information is provided.

For each setting, program repair was run $k$=\{1,2,3,4\} times on animation code that had errors. Two animation code samples had to be excluded due to compilation errors that did not allow the program repair stage to complete. Overall, 112 samples of self-debugged animation code were generated for the Repair$_{+Imgs}$ condition, and 112 samples were generated for the Repair$_{-Imgs}$ condition.

\subsection{Findings}

\begin{figure}[h]
    \centering
    \includegraphics[width=\columnwidth]{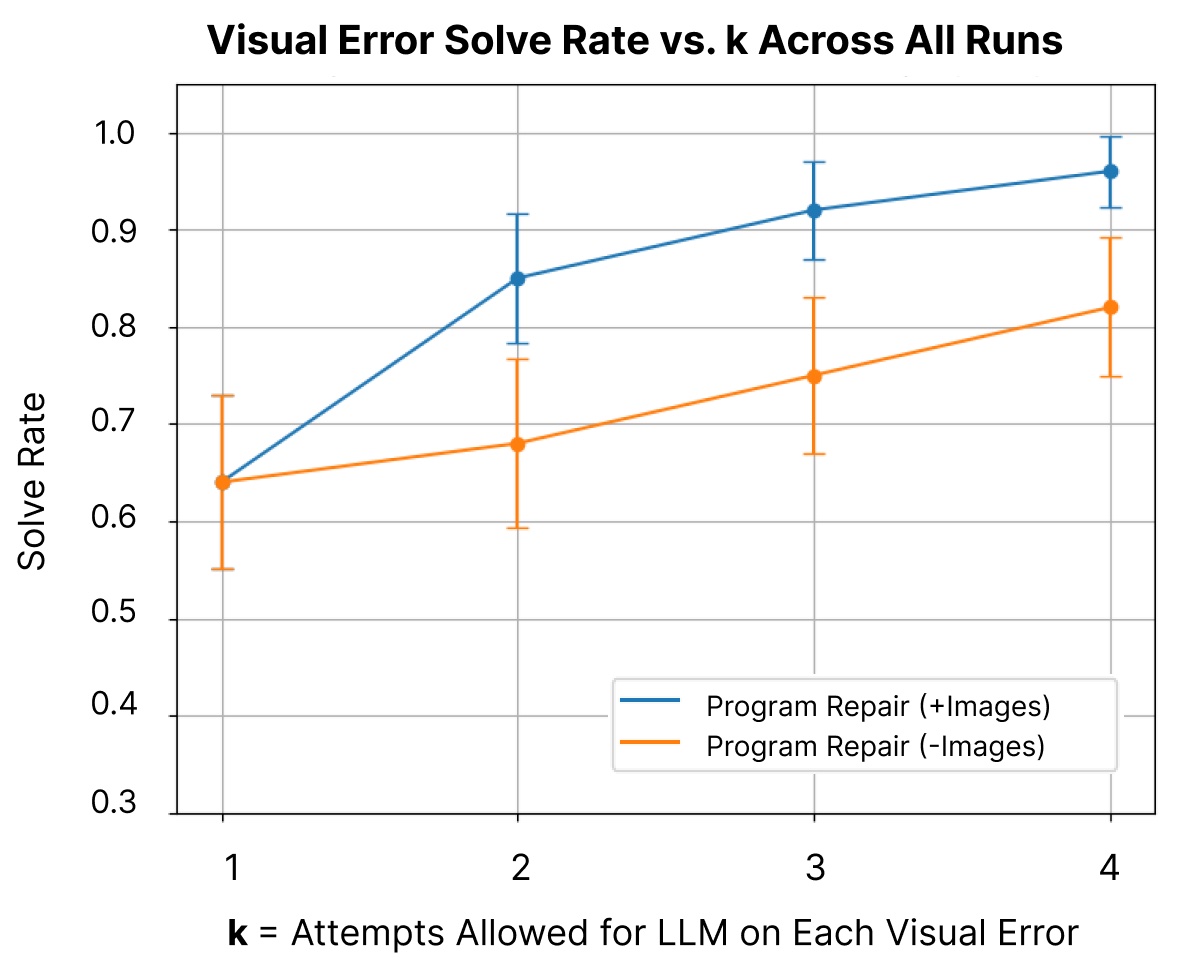}
      \caption{When LogoMotion is given more attempts to debug each error (increase in $k$), LogoMotion improves in solve rate. The trend is higher when image context is provided ($Repair_{+Imgs}$). Bars represent standard error.}
      \Description{Two lines graph the values reported in Table 2. The blue line, representing the program repair with visual context (+Images) is consistently above the orange line, which represents program repair without visual context (-Images)}
       \label{fig:solve_rate_vs_k}

\end{figure}

% \textbf{Program Repair Evaluation Graphs.} \textbf{a) } When LogoMotion is given more attempts to debug each error (increase in $k$), LogoMotion improves in solve rate. The trend is higher when image context is provided ($Repair_{+Imgs}$). Bars represent standard error. 

\begin{figure}[h]
   \centering
    \includegraphics[width=\columnwidth]{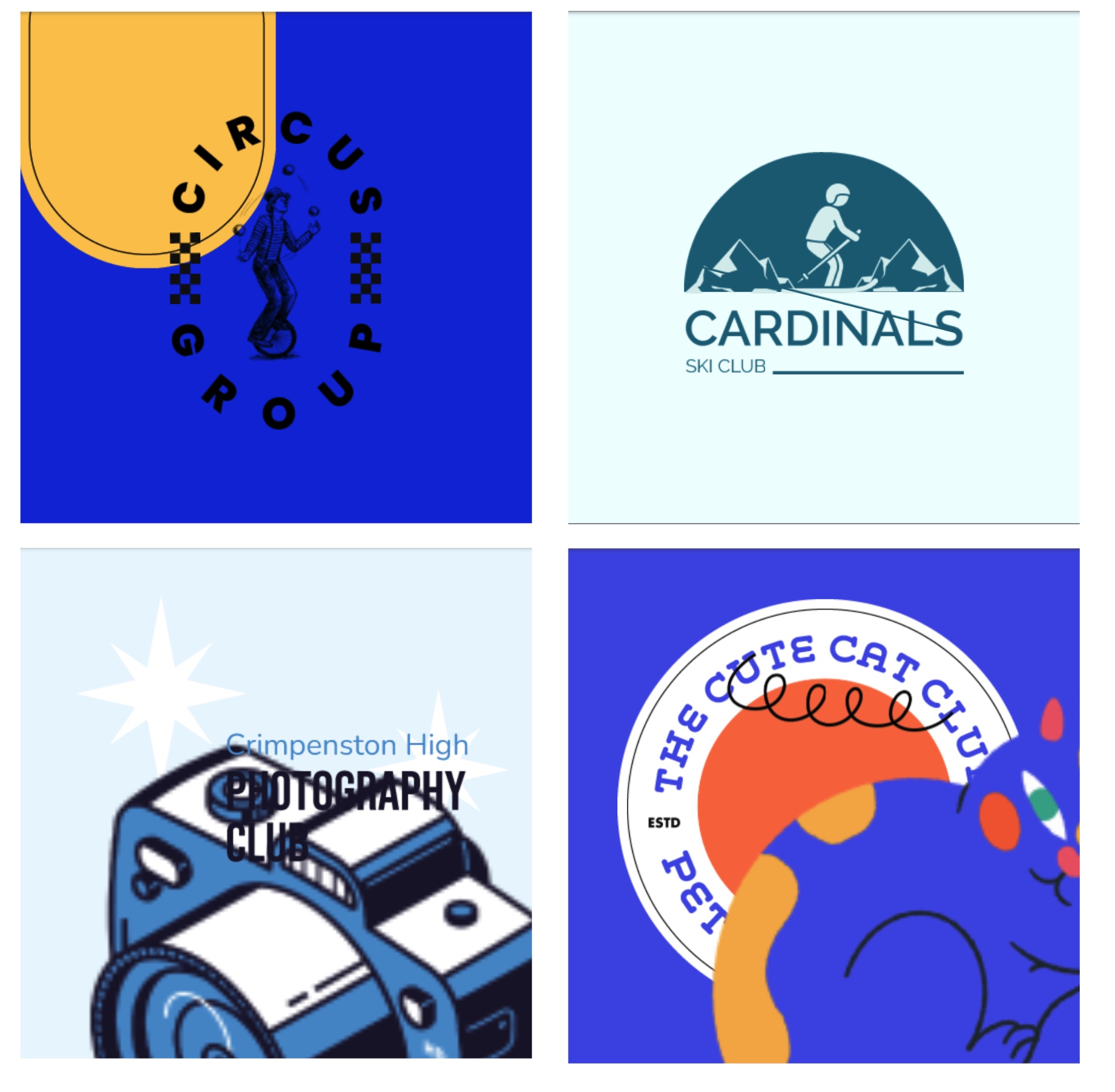}
    \caption{ Examples of animation code errors that \nickname would resolve with program repair: a position error (top left), rotation error (top right), scale errors (bottom row).}
    \label{fig:program_repair}
    
\end{figure}

\begin{table}
  \centering
  \begin{tabular}{lcc}
 k & Solve Rate$_{+I}$ & Solve Rate$_{-I}$  \\
    \hline
    1& 0.64 & 0.64    \\
    2 & 0.85 & 0.68  \\
    3& 0.92 & 0.75    \\
    4& 0.96  & 0.82
 \\
 \hline

  \end{tabular}

  \caption{ Table reporting how solve rate changes when k increases across two settings. $Solve Rate_{+I}$/$Solve Rate_{-I}$ refers to the number of runs that were error-free after $k$ attempts were made for each error $Repair_{+Imgs}$/ $Repair_{-Imgs}$ respectively.   }
  \label{tab:vary_k}
  \Description{Table showing how solve rate changes across different experimental settings for k and if image context is passed in. Solve rate improves with image context and higher k.}
\end{table}

\subsubsection{RQ3. What errors does \nickname synthesis make?}

% meaning that scale errors were less common than position errors
\nickname made 42 position-based errors in total. Position errors, which occurred when the left or top coordinate of the bounding box was off, were made in 30.4\% of the runs. \nickname made 26 scale-based errors in total, erroring in 18.4\% of the runs. These errors occurred when the width or height dimensions were off. We did not detect any opacity errors in our test set.
% could be due to the animated logo rules

Next, we provide qualitative context as to how \nickname would make these errors. Common errors resulted from not following the from-to format that is common to animation libraries (CSS and anime.js). In spite of the prompt suggesting a from-to format, keyframes would often be suggested with arrays that had over two values, so the element would not return back to its original position. For example, if the generated animation set the translateX values [10,-10, 0]--the element would end with a -10 offset relative to its correct position. Examples of errors \nickname could make are depicted in Figure \ref{fig:attempts_per_error}.

Position errors could also occur when there was an inconsistent application of absolute and relative percentages. For example, a line layer in an animation could be instructed to stretch in from 0\% outwards to 100\%. This 100 percent was intended to be with respect to the element but actually rendered with respect to the full canvas dimensions. Another frequent error was when the VLM assigned a looping animation, which would instantiate a small periodic action with the loop parameter set to true. Looping animation events generally left the elements at small deltas from their intended positions but were easily resolved.

\subsubsection{RQ4. How well can \nickname fix its errors?}

% This is pictured in Figure \ref{fig:attempts_per_error}, by the predominance of the green bar for ``Solved in 1" for each value of $k$. Note that Figure \ref{fig:solve_rate_vs_k} normalizes the number of elements, because it reports the proportion of animation code runs made error-free. Figure \ref{fig:attempts_per_error} aggregates across all errors on all design elements and reports the proportion of design elements that were made error-free within \textit{k} attempts.% This distinction is important because the one run that could not be resolved (Figure \ref{fig:solve_rate_vs_k}, k=4) had many elements whose individual errors could not be resolved (Figure\ref{fig:attempts_per_error}, in k=4).
% This is why the solve rate is different at k=4 across the graphs.
Many errors were simple enough that they would only take only one attempt from \nickname to solve. Figure \ref{fig:solve_rate_vs_k} shows that as $k$ increases, so too does the solve rate on the visual bugs for both $Repair_{+Imgs}$ and $Repair_{-Imgs}$ By $Repair_{+Imgs}$ at k=4, 96\% of the runs are resolved. There is a boost in performance from passing in visual context. Applying a Fisher's exact test (considering the small sample size of debug runs), we created a contingency table, where one variable was the presence of image context and another was success or failure at debugging, aggregated across $k$. We found that the odds ratio was 1.97 ($p = 0.08$), suggesting that the odds of successfully fixing errors with visual context nearly doubles the odds of fixing errors without visual context.

\subsubsection{Visually-Grounded Program Repair Conclusion.} Our analysis suggests that \nickname can effectively make use of the visual context it receives during program repair to self-debug. Providing the model with more attempts can improve the solve rate of \nickname on visual errors.

\section{User Study with Novices}
We conducted a user study to address the following research question: \textit{RQ5) To what extent can \nickname's code-connected widgets help novices with animation editing? }

\subsubsection{Participants}
We recruited participants through mailing lists at two universities, online communities for design tools, and from within a design company. Our target population were users who have design backgrounds and an interest in animation, but not necessarily the interest to engage with professional-grade tools. In total, sixteen participants were recruited (n=16, 6 female, 9 male, 1 non-binary, age 20-32). Fourteen participants characterized themselves as novices to animation, while two participants characterized themselves as people who had dabbled in animation as a hobby before (tried tutorials but did not master the software). Of these sixteen participants, seven had some prior experience with design. The study took 60-75 minutes for participants to complete, and participants were compensated \$20 for their time. Prior to the study, participants were sent an IRB information sheet about the experiment and a consent form.

\subsection{Experimental Design}

We conducted a within-subjects experiment and created an ablated version of our interface to use as a baseline. The baseline presented the user with only natural language interaction through prompt. We justify this as our baseline for the following reasons: 1) a prompt-only interface is one of the norms for generative interaction, 2) this baseline still had access to the same intermediate representations of the canvas (augmented HTML, generated code, logo image), 3) timelines can add overhead for novices compared to the simplicity of a prompt box, 4) \nickname's interface affordances can raise expectations even though \nickname's code-connected widgets do not behave as deterministically as traditional GUI.

Participants were given a brief introduction to the logo animation. They were instructed that their task was to animate two logo layouts given \nickname and the baseline. Prior to each task, an experimenter gave them a walkthrough of each interface with an example logo. The experimenter demonstrated the prompt box, narrative timeline, quick actions, and layer regrouping. To minimize for ordering and learning effects, the ordering of the conditions was counterbalanced and randomized across participants.

Participants picked a different layout to work with between tasks, selecting from a set of options. In total, participants animated 24 unique templates (Figure~\ref{fig:template_examples}) spanning a wide range of design elements, use cases, and layouts. To focus on the editing aspect, four logo animations for each layout were pre-generated--but not cherrypicked. Participants had 25 minutes to customize one of the four logo animations, though they could stop whenever they felt done. During the task, participants were instructed to think aloud. Between tasks, participants filled out a brief survey with Likert-scale questions. After the tasks, a semi-structured interview was conducted to understand participant experiences. Usage logs additionally captured interaction traces with both interfaces.

\subsection{Results}

Overall, many participants enjoyed working with \nickname, found the system highly performant, and saw its potential for animation.

% Independent of the condition, the majority of participants agreed that \nickname enabled them to create and edit logo animation. 
\subsubsection{Ability.} Fourteen of 16 participants using LogoMotion and 13 of 16 participants using the baseline agreed ($\geq$ 5 out of a 7-point Likert scale) that the system helped them create animations they would have otherwise not been able to create. P2 said, \textit{“It’s really cool especially for someone who couldn’t do it without this tool and wouldn’t spend a lot of time with Youtube videos or tutorials to do the very basics.”} P16 said, \textit{"I really liked it, I enjoyed it. It was really fun. I wouldn't be able to make something to this perfection if I were to make it from scratch using something like After Effects--it would have taken two hours to do it."}

Participants spent more time editing and iterating with LogoMotion than with the baseline. On average, participants spent 8.53 minutes with the baseline and 13.15 minutes with \nickname, a difference that was significant ($p = 0.017$). Participants often stopped earlier in the baseline than with \nickname because they ran out of ideas for how to edit the animation (P9), had difficulty coming up with the vocabulary to describe the animation they wanted (P1, P6, P8, P13), and because they said further iteration on timing would be frustrating (P4).

\subsubsection{Exploration}

\begin{figure}[h]
    \centering
   
        \includegraphics[width=\columnwidth]{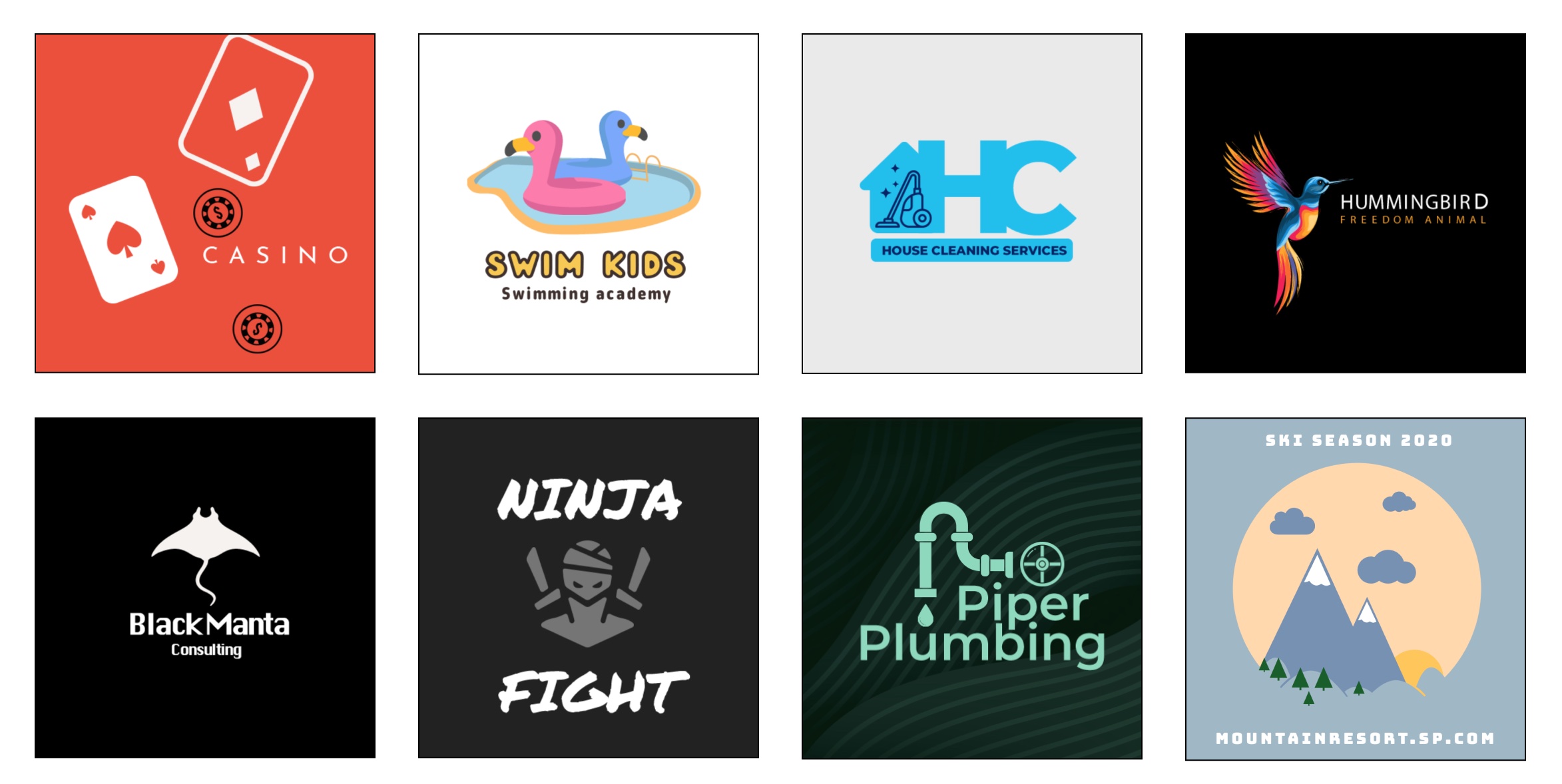}
        \caption{Examples of logos participants animated spanning a diverse range of use cases, layouts, and content. }
        \label{fig:template_examples}
        \Description{4x2 grid of templates participants animated: a casino, a SWIM KIDS swimming academy, a house cleaning ad, a hummingbird logo, a Black Manta consulting logo, Ninja Fight, Piper Plumbing logo, and Ski Season logo with a central graphic of a mountain inscribed in a circle.}

    \end{figure}

One aspect participants liked in particular was \nickname's ability to provide multiple options for inspiration. P4 was animating a Casino logo (pictured in Fig. \ref{fig:template_examples}) when they saw in the narrative timeline that one option animated the primary element to have a card flipping motion. \textit{“I like that I could see the timeline, so when I look at different options I could see what was happening under the hood. I figured out that I could use the keyword flip because some other one [animation] used the word flip.”} The narrative timeline gave novices understanding about how the system was arriving at the animation concepts and and helped users ideate new ways to animate.

\begin{figure}[h]

    \centering
    \includegraphics[width=\columnwidth]{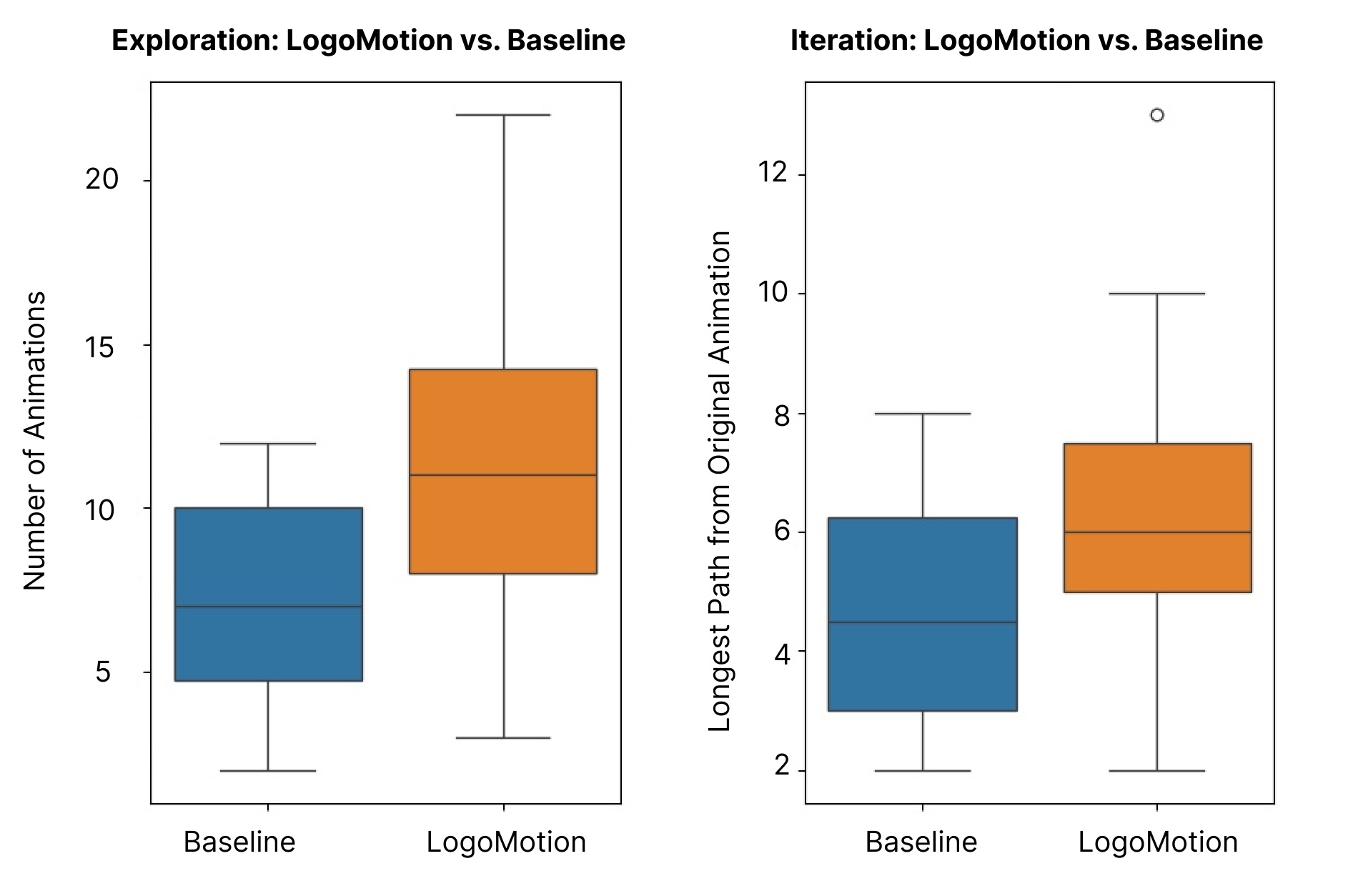}  

   \caption{ Boxplots show the distribution of animations explored and the iteration depth. Participants explored significantly more animations and had longer lengths of iteration with \nickname over the baseline.}
   \Description{Boxplots for LogoMotion. Exploration: LogoMotion vs. Baseline. Iteration: LogoMotion vs. Baseline. For Exploration, LogoMotion has a higher median and maximum for number of animations. For Iteration, LogoMotion has a higher median and a higher maximum for length of animations (an outlier). }
    \label{fig:explorationtree}
    % \label{fig:explorationtree}
 
\end{figure}

The version history captured by the Exploration Tree allowed us to measure exploration and iteration. For each participant, we measured the number of edits made and depth of iteration (longest path from base animation to last edited animation). Participants explored significantly more animations with \nickname compared to the baseline, averaging 11.19 animations with \nickname versus 7.13 animations with the baseline ($p \leq 0.01$). Participants were also able to carry out more rounds of iteration with \nickname, creating deeper chains of edits. On average, the maximum depth of iteration in the full condition was 6.56 animation edits in length compared to 4.75 animation edits in length, a difference that was statistically significant ($p \leq$ 0.05). We visualize the distribution of animations explored and iteration depth across participants in Fig. \ref{fig:explorationtree}.

\begin{figure}[h]

   \centering
        \includegraphics[width=\columnwidth]{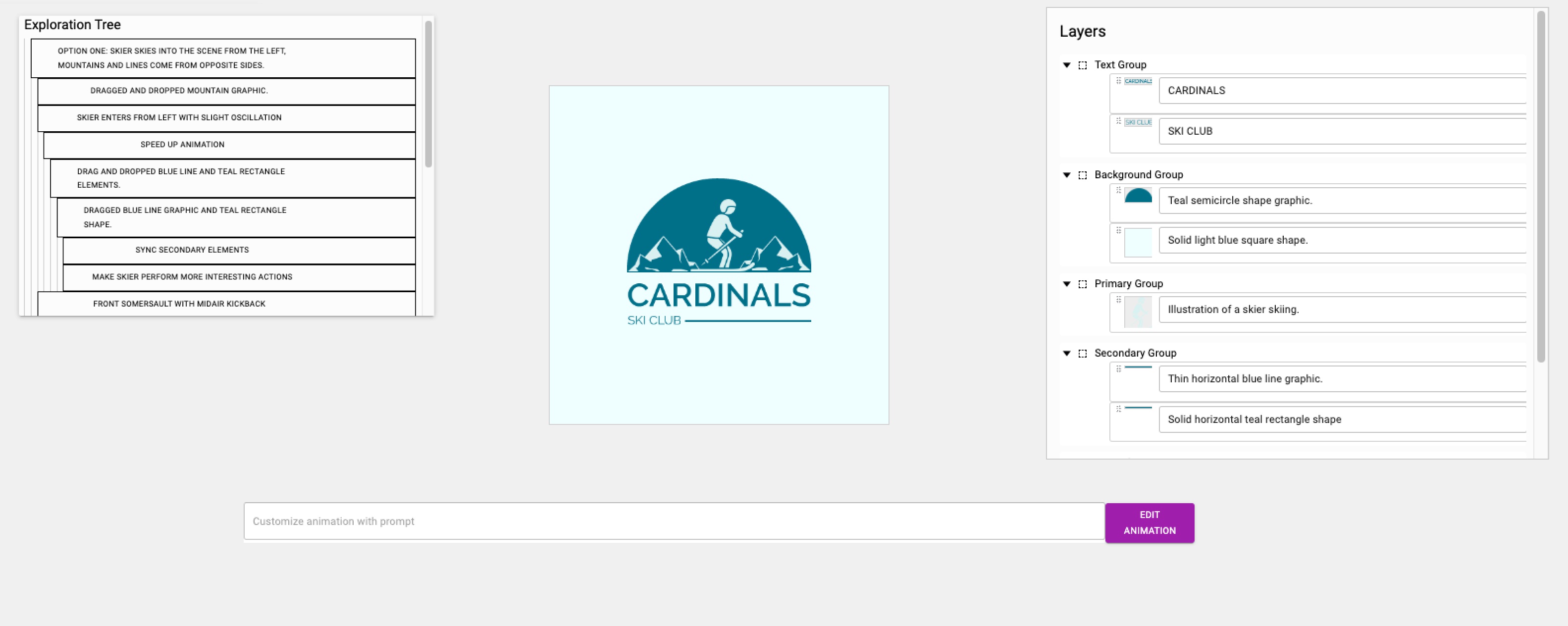}
        \caption{Our baseline condition included the Exploration Tree, Prompt Box, and Layer panel. It provided prompting as the main form of interaction.}
        \label{fig:baseline}
        \Description{Baseline user interface is depicted. Top left shows the Exploration Tree, a tree view of generated animations. The layers panel in top right shows layer information like captions and groups LogoMotion assigns. Below the canvas, there is a textbox for a user's prompt. }
\end{figure}

\subsubsection{Types of Customization.}

Participants were expressive in their natural language editing requests. P14 edited a logo featuring a manta with the prompt: \textit{``During hero moment, stingray floats backward as if fighting against a current"}, going for an interaction between the primary element and background. In animating a logo for Piper Plumbing, P1 wrote (\textit{``water droplet appears as if it is an actual drop of water coming from the pipeline, after the wheel has turned once"}) and was pleased to see that \nickname could fulfill such a complex edit. P16 iterated on getting a primary logo element to have a ``gentle baby walk" entrance into the frame using both prompts and the ``Subtle" quick action. 

\section{Discussion}

We have introduced \nickname, VLM-powered code generation approach that helps users create semantically meaningful animation using 1) visually-grounded code synthesis and repair and 2) code-connected editing widgets. We now discuss how the merits of these contributions and how they generalize beyond logo animation. 

\subsection{Generalizability of Visually-Grounded Code Synthesis and Repair}

Visually-grounded code synthesis is powerful because it unlocks how code generation approaches can be guided by an AI's ability to analyze images. \nickname shows how to do this by building an HTML representation that is visually aware of what is on the canvas. The HTML representation reflects the visual hierarchy of the elements in the image and captures grouping information in the code hierarchy organization. \nickname demonstrates how a VLM can be applied to understand layer content, automatically make groups, and update the HTML representation when users interact with a layers panel. This is relevant to any design tool that has a layer panel. In tools like Adobe Illustrator and Figma, users engage with hundreds of layers and groups, and \nickname shows how an AI approach can solve classic painpoints like layer organization and group layers for users. Secondly, \nickname shows how a code generation approach can be guided by a design concept. This is powerful because it allows animation to move beyond templates and presets --\nickname outperforms an industry approach Magic Animate in this respect--and move towards semantically meaningful animation. Visually-grounded code synthesis can also extend to other animation tasks like slide animation (where visual hierarchy and grouping is key) or kinetic typography (where the motion must match the semantic meaning of the text). 

% UI/UX prototyping, which is a domain where visuals and code are tightly connected, 
% This representation can generalize to many other animation tasks such as slide animation, or kinetic typography, where there is also a semantic meaning that the animation has to reinforce.  

% Furthermore, its visual grounding steps can be expanded to support other important kinds of visual understanding involving grids, patterns, and gestalt. HTML representation informed by visual analysis 

% If an AI generates a video of a falling apple, can the motion be visually checked to see if the video reflects the correct physics? 
\nickname's visually-grounded program repair mechanism also has high generalizability. It can be effective for many other tasks where code renders visual output such as UI mockups, front-end prototyping, and game design. For example, if a UI prototype is generated from a visual mockup, it can be checked for if whether components meet the visual specification of the mockup. Furthermore, for this checking process to be able to isolate visual differences, visual reflection has to be able to support different ``beams of focus". \nickname shows how to implement visually-grounded program repair with a layer-wise beam of focus and effectively solve 96\% of detected errors within four attempts. Though in \nickname we only checked the last frame of the animation against the target layout, in future work, we can check intermediate frames of animation to build a deeper understanding of motion.

% Other beams of focus could include components within a UI prototype or objects within a 3D scene. 

\subsection{Generalizability of Code-Connected AI Editing Widgets}

Code-connected AI editing widgets allow users to enjoy the benefits of a code representation (controllability, greater expressive range) without having to engage with the technical details of implementation. \nickname's narrative timeline widget allowed users to pair the expression gained from prompts with the other modes of controls that come from GUIs (e.g. selection, reordering, drag-and-drop). Each time the animation was edited, the AI implementation would automatically produce a well-eased and smooth animation, without requiring any tedious manipulation of timestamps or fine-grained editing that is common with traditional tools. The narrative timeline that we introduce is also a novel improvement upon traditional timelines, which are not capable of describing the events that take place within their keyframes and blocks. Narrative timelines can generalize to other design tasks such as video editing. Overall, these code-connected AI widgets are new components that open up interactions beyond the standard chatbot paradigm.

\subsubsection{Semantic vs. Specific Controls for Animation Editing}

\nickname allowed us to explore when it is best to support users with generative, semantic-based controls (prompts), when it is best to support users with granular specific controls (timelines), and when there are opportunities to bind these two types of controls together within an interface. The obvious benefit of semantic controls is that it opens up animation to novices--expressing animation is as simple as expressing a thought or story. Furthermore, in language it is easier to have conceptual reach and to divergently explore animations. Participants used properties of language like metaphors (\textit{``gentle baby walk"}) and action description (\textit{``swimming as if fighting against a current"}) to create connections that reinforced the high-level visual message. By introducing a design concept to have semantic control over, \nickname helps users engage with code at the level of narrative. However, language as the sole mode of expression is insufficient--it lacks the control necessary for animation, which has many dimensions that rely on perceptual feel such as timing, dynamics, synchronization, grouping, and interactions between elements. These aspects are hard for novices to find the language for; a novice can write that two objects should overlap in timing, but overlap by how much, and which element should come first? If something should be faster, by how much? What does it mean for two elements to ``come in together"? Quickly, it becomes unwieldy to find the right balance between underspecification and overspecification in natural language. \nickname's code-connected components allowed users to divert many of these editing goals to GUI interactions that are second nature (e.g. deterministically setting the timing on elements, or how elements group in the organization of the animation). At the same time, \nickname demonstrated how semantic and specific controls can be paired together to regenerate the underlying code. Pairing prompts to blocks on the narrative timeline by selection helped ground semantic intents in aspects of the code, supporting local exploration with constraints such as what elements were being targeted or what snippet of the underlying code should be edited. Overall, \nickname showed how semantic and specific controls can be integrated within an interface to span the different levels of abstraction necessary for animation creation and iteration.

\subsection{Limitations and Future Work}

\revAdd{We discussed \nickname outputs with professional motion designers, who provided insights about \nickname's limitations. While  of quality, \nickname does not produce animations that are on a professional level. The animations do not at a meta-level guide the viewer's eye. The animations also often unevenly apply fundamental principles of animation. They mentioned how the the design concept could excessively apply overlapping action and could miss the opportunity to create complex interactions between groups. \nickname also does not support complex motion paths, morphing, and physics-based animation. We made efforts to explore path motion but found that it was hard to integrate an SVG path defined outside of the design concept into the generated code. Motion paths that can define primary and secondary motion, incorporate richer curved motion, and open up the opportunity for direct manipulation represent an important next step. } 

\revAdd{Another limitation is that \nickname scoped around logos. To focus on creating semantically meaningful animation, we made design decisions specific to logos (e.g. there is generally one hero element that should get more emphasis in the animation). However, other formats have different motion design principles. We experimented with having \nickname animate social media posts as well as a few webpages. We found that in these formats where there is more text, the animation had to be more subtle to balance readability and advertisement. This helped further establish the importance of having layer panel interactions and quick actions (Synchronize, Make Subtle). To adapt to other domains like CSS animation or vector animation, \nickname would likely have to adapt to generate and check on color and shape change. It is possible that the focus on the entrance description and use of the word ``motion" predisposed \nickname to animate in terms of motion rather than these other properties. }

\revAdd{Lastly in terms of our user evaluation, we only tested \nickname with sixteen people. For three of the early participants, the drag-and-drop interaction was too sensitive, so accidental interactions had to be excluded from analysis. Another limitation was that some participants had less fruitful experiences with \nickname because logo they chose to work with did not present many opportunities for animation. The logo elements, being all raster images, did not support deformable motion and could not be broken further down into parts. \nickname ideally could have given feedback for what edits would be more possible to fulfill and what was out of bounds. 
 }

% Our program repair stage could have also been a larger-scale evaluation over more datapoints.

\section{Conclusion}
\nickname presents an VLM-powered tool that allows users to automatically and interactively animate logos and create semantically meaningful motion. Its approach has three stages:  visually-grounded code synthesis, visually-grounded program repair, and code-connected widgets. Within this editor, a narrative timeline, layer panel, and quick actions helped users edit for motion, grouping, and timing of their animation. We show in evaluations that \nickname outperforms a state-of-the-art industry tool at producing animation that is relevant to its design elements. \nickname was found to be expressive and enjoyable for novices to create and edit logo animation.

\bibliographystyle{ACM-Reference-Format}
\bibliography{citations}


\begin{thebibliography}{54}

%%% the Web address.

\ifx \showCODEN    \undefined \def \showCODEN     #1{\unskip}     \fi
\ifx \showDOI      \undefined \def \showDOI       #1{#1}\fi
\ifx \showISBNx    \undefined \def \showISBNx     #1{\unskip}     \fi
\ifx \showISBNxiii \undefined \def \showISBNxiii  #1{\unskip}     \fi
\ifx \showISSN     \undefined \def \showISSN      #1{\unskip}     \fi
\ifx \showLCCN     \undefined \def \showLCCN      #1{\unskip}     \fi
\ifx \shownote     \undefined \def \shownote      #1{#1}          \fi
\ifx \showarticletitle \undefined \def \showarticletitle #1{#1}   \fi
\ifx \showURL      \undefined \def \showURL       {\relax}        \fi
% invisible to TeX
\providecommand\bibfield[2]{#2}
\providecommand\bibinfo[2]{#2}
\providecommand\natexlab[1]{#1}
\providecommand\showeprint[2][]{arXiv:#2}

\bibitem[ble(2023)]%
        {blendergpt}
 \bibinfo{year}{2023}\natexlab{}.
\newblock \bibinfo{title}{BlenderGPT}.
\newblock
\newblock
\urldef\tempurl%
\url{https://github.com/gd3kr/BlenderGPT}
\showURL{%
\tempurl}


\bibitem[Cap(2023)]%
        {Capcut}
 \bibinfo{year}{2023}\natexlab{}.
\newblock \bibinfo{title}{Capcut Templates}.
\newblock
\newblock
\urldef\tempurl%
\url{https://www.capcut.com/templates}
\showURL{%
\tempurl}


\bibitem[Ado(2023)]%
        {AdobePremiereTemplates}
 \bibinfo{year}{2023}\natexlab{}.
\newblock \bibinfo{title}{Motion Array Premiere Templates}.
\newblock
\newblock
\urldef\tempurl%
\url{https://motionarray.com/browse/premiere-pro-templates/}
\showURL{%
\tempurl}


\bibitem[Pin(2023)]%
        {PinterestShuffles}
 \bibinfo{year}{2023}\natexlab{}.
\newblock \bibinfo{title}{Pinterest Shuffles}.
\newblock
\newblock
\urldef\tempurl%
\url{https://www.shffls.com/}
\showURL{%
\tempurl}


\bibitem[Ado(2024)]%
        {AdobeExpress:AnimatedTemplates}
 \bibinfo{year}{2024}\natexlab{}.
\newblock \bibinfo{title}{Adobe Express Animated Templates}.
\newblock
\newblock
\urldef\tempurl%
\url{https://www.adobe.com/express/create/logo/animated}
\showURL{%
\tempurl}


\bibitem[Can(2024)]%
        {Canva:MagicAnimate}
 \bibinfo{year}{2024}\natexlab{}.
\newblock \bibinfo{title}{Canva Magic Animate}.
\newblock
\newblock
\urldef\tempurl%
\url{https://www.canva.com/help/using-magic-animate/}
\showURL{%
\tempurl}


\bibitem[Fig(2024)]%
        {Figma:SmartAnimate}
 \bibinfo{year}{2024}\natexlab{}.
\newblock \bibinfo{title}{Figma Smart Animate}.
\newblock
\newblock
\urldef\tempurl%
\url{https://help.figma.com/hc/en-us/articles/360039818874-Smart-animate-layers-between-frames}
\showURL{%
\tempurl}


\bibitem[Pow(2024)]%
        {Powerpoint:Morph}
 \bibinfo{year}{2024}\natexlab{}.
\newblock \bibinfo{title}{Powerpoint Morph}.
\newblock
\newblock
\urldef\tempurl%
\url{https://support.microsoft.com/en-us/office/use-the-morph-transition-in-powerpoint}
\showURL{%
\tempurl}


\bibitem[Run(2024)]%
        {Runway:MotionBrush}
 \bibinfo{year}{2024}\natexlab{}.
\newblock \bibinfo{title}{Runway Motion Brush}.
\newblock
\newblock
\urldef\tempurl%
\url{https://academy.runwayml.com/gen2/gen2-motion-brush}
\showURL{%
\tempurl}


\bibitem[Adobe(2022)]%
        {intro_looping_outro}
\bibfield{author}{\bibinfo{person}{Adobe}.} \bibinfo{year}{2022}\natexlab{}.
\newblock \bibinfo{title}{Animate text, videos and photos in Adobe Express.}
\newblock
\newblock
\urldef\tempurl%
\url{https://helpx.adobe.com/express/using/animation.html}
\showURL{%
\tempurl}


\bibitem[Angert et~al\mbox{.}(2023a)]%
        {Angert:23:Spellburst}
\bibfield{author}{\bibinfo{person}{Tyler Angert}, \bibinfo{person}{Miroslav Suzara}, \bibinfo{person}{Jenny Han}, \bibinfo{person}{Christopher Pondoc}, {and} \bibinfo{person}{Hariharan Subramonyam}.} \bibinfo{year}{2023}\natexlab{a}.
\newblock \showarticletitle{Spellburst: A Node-based Interface for Exploratory Creative Coding with Natural Language Prompts}. In \bibinfo{booktitle}{\emph{Proceedings of the 36th Annual ACM Symposium on User Interface Software and Technology}} \emph{(\bibinfo{series}{UIST ’23})}. \bibinfo{publisher}{ACM}, \bibinfo{pages}{1–22}.
\newblock
\urldef\tempurl%
\url{https://doi.org/10.1145/3586183.3606719}
\showDOI{\tempurl}


\bibitem[Angert et~al\mbox{.}(2023b)]%
        {spellburst}
\bibfield{author}{\bibinfo{person}{Tyler Angert}, \bibinfo{person}{Miroslav Suzara}, \bibinfo{person}{Jenny Han}, \bibinfo{person}{Christopher Pondoc}, {and} \bibinfo{person}{Hariharan Subramonyam}.} \bibinfo{year}{2023}\natexlab{b}.
\newblock \showarticletitle{Spellburst: A Node-based Interface for Exploratory Creative Coding with Natural Language Prompts}. In \bibinfo{booktitle}{\emph{Proceedings of the 36th Annual ACM Symposium on User Interface Software and Technology}} \emph{(\bibinfo{series}{UIST ’23})}. \bibinfo{publisher}{ACM}.
\newblock
\urldef\tempurl%
\url{https://doi.org/10.1145/3586183.3606719}
\showDOI{\tempurl}


\bibitem[Bang et~al\mbox{.}(2023)]%
        {Bang:2023:Multitask}
\bibfield{author}{\bibinfo{person}{Yejin Bang}, \bibinfo{person}{Samuel Cahyawijaya}, \bibinfo{person}{Nayeon Lee}, \bibinfo{person}{Wenliang Dai}, \bibinfo{person}{Dan Su}, \bibinfo{person}{Bryan Wilie}, \bibinfo{person}{Holy Lovenia}, \bibinfo{person}{Ziwei Ji}, \bibinfo{person}{Tiezheng Yu}, \bibinfo{person}{Willy Chung}, \bibinfo{person}{Quyet~V. Do}, \bibinfo{person}{Yan Xu}, {and} \bibinfo{person}{Pascale Fung}.} \bibinfo{year}{2023}\natexlab{}.
\newblock \bibinfo{title}{A Multitask, Multilingual, Multimodal Evaluation of ChatGPT on Reasoning, Hallucination, and Interactivity}.
\newblock
\newblock
\showeprint[arxiv]{2302.04023}~[cs.CL]
\urldef\tempurl%
\url{https://arxiv.org/abs/2302.04023}
\showURL{%
\tempurl}


\bibitem[Cao et~al\mbox{.}(2023)]%
        {Cao:23:DataParticles}
\bibfield{author}{\bibinfo{person}{Yining Cao}, \bibinfo{person}{Jane~L E}, \bibinfo{person}{Zhutian Chen}, {and} \bibinfo{person}{Haijun Xia}.} \bibinfo{year}{2023}\natexlab{}.
\newblock \showarticletitle{DataParticles: Block-based and Language-oriented Authoring of Animated Unit Visualizations}. In \bibinfo{booktitle}{\emph{Proceedings of the 2023 CHI Conference on Human Factors in Computing Systems}} (, Hamburg, Germany,) \emph{(\bibinfo{series}{CHI '23})}. \bibinfo{publisher}{Association for Computing Machinery}, \bibinfo{address}{New York, NY, USA}, Article \bibinfo{articleno}{808}, \bibinfo{numpages}{15}~pages.
\newblock
\showISBNx{9781450394215}
\urldef\tempurl%
\url{https://doi.org/10.1145/3544548.3581472}
\showDOI{\tempurl}


\bibitem[Chen et~al\mbox{.}(2023a)]%
        {chen2023selfdebug}
\bibfield{author}{\bibinfo{person}{Xinyun Chen}, \bibinfo{person}{Maxwell Lin}, \bibinfo{person}{Nathanael Schärli}, {and} \bibinfo{person}{Denny Zhou}.} \bibinfo{year}{2023}\natexlab{a}.
\newblock \bibinfo{title}{Teaching Large Language Models to Self-Debug}.
\newblock
\newblock
\showeprint[arxiv]{2304.05128}~[cs.CL]


\bibitem[Chen et~al\mbox{.}(2023b)]%
        {Chen:2023:Seine}
\bibfield{author}{\bibinfo{person}{Xinyuan Chen}, \bibinfo{person}{Yaohui Wang}, \bibinfo{person}{Lingjun Zhang}, \bibinfo{person}{Shaobin Zhuang}, \bibinfo{person}{Xin Ma}, \bibinfo{person}{Jiashuo Yu}, \bibinfo{person}{Yali Wang}, \bibinfo{person}{Dahua Lin}, \bibinfo{person}{Yu Qiao}, {and} \bibinfo{person}{Ziwei Liu}.} \bibinfo{year}{2023}\natexlab{b}.
\newblock \bibinfo{title}{SEINE: Short-to-Long Video Diffusion Model for Generative Transition and Prediction}.
\newblock
\newblock
\showeprint[arxiv]{2310.20700}~[cs.CV]


\bibitem[Cosler et~al\mbox{.}(2023)]%
        {nl2spec}
\bibfield{author}{\bibinfo{person}{Matthias Cosler}, \bibinfo{person}{Christopher Hahn}, \bibinfo{person}{Daniel Mendoza}, \bibinfo{person}{Frederik Schmitt}, {and} \bibinfo{person}{Caroline Trippel}.} \bibinfo{year}{2023}\natexlab{}.
\newblock \bibinfo{title}{nl2spec: Interactively Translating Unstructured Natural Language to Temporal Logics with Large Language Models}.
\newblock
\newblock
\showeprint[arxiv]{2303.04864}~[cs.LO]


\bibitem[Crowson({[n.\,d.]})]%
        {discodiffusion}
\bibfield{author}{\bibinfo{person}{Katherine Crowson}.} \bibinfo{year}{[n.\,d.]}\natexlab{}.
\newblock \bibinfo{title}{Alembics/disco-diffusion}.
\newblock
\newblock
\urldef\tempurl%
\url{https://github.com/alembics/disco-diffusion}
\showURL{%
\tempurl}


\bibitem[Davis et~al\mbox{.}(2008)]%
        {KSketch}
\bibfield{author}{\bibinfo{person}{Richard~C. Davis}, \bibinfo{person}{Brien Colwell}, {and} \bibinfo{person}{James~A. Landay}.} \bibinfo{year}{2008}\natexlab{}.
\newblock \showarticletitle{K-sketch: a 'kinetic' sketch pad for novice animators}. In \bibinfo{booktitle}{\emph{Proceedings of the SIGCHI Conference on Human Factors in Computing Systems}} (Florence, Italy) \emph{(\bibinfo{series}{CHI '08})}. \bibinfo{publisher}{Association for Computing Machinery}, \bibinfo{address}{New York, NY, USA}, \bibinfo{pages}{413–422}.
\newblock
\showISBNx{9781605580111}
\urldef\tempurl%
\url{https://doi.org/10.1145/1357054.1357122}
\showDOI{\tempurl}


\bibitem[Dontcheva et~al\mbox{.}(2003)]%
        {Dontcheva:03:Layered}
\bibfield{author}{\bibinfo{person}{Mira Dontcheva}, \bibinfo{person}{Gary Yngve}, {and} \bibinfo{person}{Zoran Popovi\'{c}}.} \bibinfo{year}{2003}\natexlab{}.
\newblock \showarticletitle{Layered acting for character animation}.
\newblock \bibinfo{journal}{\emph{ACM Trans. Graph.}} \bibinfo{volume}{22}, \bibinfo{number}{3} (\bibinfo{date}{July} \bibinfo{year}{2003}), \bibinfo{pages}{409–416}.
\newblock
\showISSN{0730-0301}
\urldef\tempurl%
\url{https://doi.org/10.1145/882262.882285}
\showDOI{\tempurl}


\bibitem[Duan et~al\mbox{.}(2024)]%
        {automaticfeedback}
\bibfield{author}{\bibinfo{person}{Peitong Duan}, \bibinfo{person}{Jeremy Warner}, \bibinfo{person}{Yang Li}, {and} \bibinfo{person}{Bjoern Hartmann}.} \bibinfo{year}{2024}\natexlab{}.
\newblock \showarticletitle{Generating Automatic Feedback on UI Mockups with Large Language Models}. In \bibinfo{booktitle}{\emph{Proceedings of the CHI Conference on Human Factors in Computing Systems}} \emph{(\bibinfo{series}{CHI ’24}, Vol.~\bibinfo{volume}{4})}. \bibinfo{publisher}{ACM}, \bibinfo{pages}{1–20}.
\newblock
\urldef\tempurl%
\url{https://doi.org/10.1145/3613904.3642782}
\showDOI{\tempurl}


\bibitem[et. al.(2021)]%
        {humaneval}
\bibfield{author}{\bibinfo{person}{Mark~Chen et. al.}} \bibinfo{year}{2021}\natexlab{}.
\newblock \bibinfo{title}{Evaluating Large Language Models Trained on Code}.
\newblock
\newblock
\showeprint[arxiv]{2107.03374}~[cs.LG]


\bibitem[et. al.(2024)]%
        {gpt4}
\bibfield{author}{\bibinfo{person}{OpenAI et. al.}} \bibinfo{year}{2024}\natexlab{}.
\newblock \bibinfo{title}{GPT-4 Technical Report}.
\newblock
\newblock
\showeprint[arxiv]{2303.08774}~[cs.CL]


\bibitem[Favero et~al\mbox{.}(2024)]%
        {Favero:2024:Multimodal}
\bibfield{author}{\bibinfo{person}{Alessandro Favero}, \bibinfo{person}{Luca Zancato}, \bibinfo{person}{Matthew Trager}, \bibinfo{person}{Siddharth Choudhary}, \bibinfo{person}{Pramuditha Perera}, \bibinfo{person}{Alessandro Achille}, \bibinfo{person}{Ashwin Swaminathan}, {and} \bibinfo{person}{Stefano Soatto}.} \bibinfo{year}{2024}\natexlab{}.
\newblock \bibinfo{title}{Multi-Modal Hallucination Control by Visual Information Grounding}.
\newblock
\newblock
\showeprint[arxiv]{2403.14003}~[cs.CV]
\urldef\tempurl%
\url{https://arxiv.org/abs/2403.14003}
\showURL{%
\tempurl}


\bibitem[Fiannaca et~al\mbox{.}(2023)]%
        {fiannaca23prompting}
\bibfield{author}{\bibinfo{person}{Alexander~J. Fiannaca}, \bibinfo{person}{Chinmay Kulkarni}, \bibinfo{person}{Carrie~J Cai}, {and} \bibinfo{person}{Michael Terry}.} \bibinfo{year}{2023}\natexlab{}.
\newblock \showarticletitle{Programming without a Programming Language: Challenges and Opportunities for Designing Developer Tools for Prompt Programming}. In \bibinfo{booktitle}{\emph{Extended Abstracts of the 2023 CHI Conference on Human Factors in Computing Systems}} (<conf-loc>, <city>Hamburg</city>, <country>Germany</country>, </conf-loc>) \emph{(\bibinfo{series}{CHI EA '23})}. \bibinfo{publisher}{Association for Computing Machinery}, \bibinfo{address}{New York, NY, USA}, Article \bibinfo{articleno}{235}, \bibinfo{numpages}{7}~pages.
\newblock
\showISBNx{9781450394222}
\urldef\tempurl%
\url{https://doi.org/10.1145/3544549.3585737}
\showDOI{\tempurl}


\bibitem[Gulwani et~al\mbox{.}(2017)]%
        {msr}
\bibfield{author}{\bibinfo{person}{Sumit Gulwani}, \bibinfo{person}{Alex Polozov}, {and} \bibinfo{person}{Rishabh Singh}.} \bibinfo{year}{2017}\natexlab{}.
\newblock \bibinfo{booktitle}{\emph{Program Synthesis}}. Vol.~\bibinfo{volume}{4}.
\newblock \bibinfo{publisher}{NOW}. 1--119 pages.
\newblock
\urldef\tempurl%
\url{https://www.microsoft.com/en-us/research/publication/program-synthesis/}
\showURL{%
\tempurl}


\bibitem[Gunturu et~al\mbox{.}(2024)]%
        {Gunturu:24:AugmentedPhysics}
\bibfield{author}{\bibinfo{person}{Aditya Gunturu}, \bibinfo{person}{Yi Wen}, \bibinfo{person}{Nandi Zhang}, \bibinfo{person}{Jarin Thundathil}, \bibinfo{person}{Rubaiat~Habib Kazi}, {and} \bibinfo{person}{Ryo Suzuki}.} \bibinfo{year}{2024}\natexlab{}.
\newblock \showarticletitle{Augmented Physics: Creating Interactive and Embedded Physics Simulations from Static Textbook Diagrams}. In \bibinfo{booktitle}{\emph{Proceedings of the 37th Annual ACM Symposium on User Interface Software and Technology}} (Pittsburgh, PA, USA) \emph{(\bibinfo{series}{UIST '24})}. \bibinfo{publisher}{Association for Computing Machinery}, \bibinfo{address}{New York, NY, USA}, Article \bibinfo{articleno}{144}, \bibinfo{numpages}{12}~pages.
\newblock
\showISBNx{9798400706288}
\urldef\tempurl%
\url{https://doi.org/10.1145/3654777.3676392}
\showDOI{\tempurl}


\bibitem[Guo et~al\mbox{.}(2023)]%
        {Guo:2023:Animatediff}
\bibfield{author}{\bibinfo{person}{Yuwei Guo}, \bibinfo{person}{Ceyuan Yang}, \bibinfo{person}{Anyi Rao}, \bibinfo{person}{Yaohui Wang}, \bibinfo{person}{Yu Qiao}, \bibinfo{person}{Dahua Lin}, {and} \bibinfo{person}{Bo Dai}.} \bibinfo{year}{2023}\natexlab{}.
\newblock \bibinfo{title}{AnimateDiff: Animate Your Personalized Text-to-Image Diffusion Models without Specific Tuning}.
\newblock
\newblock
\showeprint[arxiv]{2307.04725}~[cs.CV]


\bibitem[Hendrycks et~al\mbox{.}(2021)]%
        {apps}
\bibfield{author}{\bibinfo{person}{Dan Hendrycks}, \bibinfo{person}{Steven Basart}, \bibinfo{person}{Saurav Kadavath}, \bibinfo{person}{Mantas Mazeika}, \bibinfo{person}{Akul Arora}, \bibinfo{person}{Ethan Guo}, \bibinfo{person}{Collin Burns}, \bibinfo{person}{Samir Puranik}, \bibinfo{person}{Horace He}, \bibinfo{person}{Dawn Song}, {and} \bibinfo{person}{Jacob Steinhardt}.} \bibinfo{year}{2021}\natexlab{}.
\newblock \bibinfo{title}{Measuring Coding Challenge Competence With APPS}.
\newblock
\newblock
\showeprint[arxiv]{2105.09938}~[cs.SE]


\bibitem[Huang et~al\mbox{.}(2023)]%
        {llmr}
\bibfield{author}{\bibinfo{person}{Han Huang}, \bibinfo{person}{Fernanda De~La~Torre}, \bibinfo{person}{Cathy~Mengying Fang}, \bibinfo{person}{Andrzej Banburski-Fahey}, \bibinfo{person}{Judith Amores}, {and} \bibinfo{person}{Jaron Lanier}.} \bibinfo{year}{2023}\natexlab{}.
\newblock \showarticletitle{Real-time Animation Generation and Control on Rigged Models via Large Language Models}.
\newblock \bibinfo{journal}{\emph{arXiv preprint arXiv:2310.17838}} (\bibinfo{year}{2023}).
\newblock


\bibitem[Igarashi et~al\mbox{.}(2005)]%
        {Igarashi:05:AsRigid}
\bibfield{author}{\bibinfo{person}{Takeo Igarashi}, \bibinfo{person}{Tomer Moscovich}, {and} \bibinfo{person}{John~F. Hughes}.} \bibinfo{year}{2005}\natexlab{}.
\newblock \showarticletitle{As-rigid-as-possible shape manipulation}.
\newblock \bibinfo{journal}{\emph{ACM Trans. Graph.}} \bibinfo{volume}{24}, \bibinfo{number}{3} (\bibinfo{date}{July} \bibinfo{year}{2005}), \bibinfo{pages}{1134–1141}.
\newblock
\showISSN{0730-0301}
\urldef\tempurl%
\url{https://doi.org/10.1145/1073204.1073323}
\showDOI{\tempurl}


\bibitem[Jahanlou and Chilana(2022)]%
        {Jahanlou:22:Katika}
\bibfield{author}{\bibinfo{person}{Amir Jahanlou} {and} \bibinfo{person}{Parmit~K Chilana}.} \bibinfo{year}{2022}\natexlab{}.
\newblock \showarticletitle{Katika: An End-to-End System for Authoring Amateur Explainer Motion Graphics Videos}. In \bibinfo{booktitle}{\emph{Proceedings of the 2022 CHI Conference on Human Factors in Computing Systems}} (New Orleans, LA, USA) \emph{(\bibinfo{series}{CHI '22})}. \bibinfo{publisher}{Association for Computing Machinery}, \bibinfo{address}{New York, NY, USA}, Article \bibinfo{articleno}{502}, \bibinfo{numpages}{14}~pages.
\newblock
\showISBNx{9781450391573}
\urldef\tempurl%
\url{https://doi.org/10.1145/3491102.3517741}
\showDOI{\tempurl}


\bibitem[Johnston and Thomas(1981)]%
        {Disney:81:PrinciplesAnimation}
\bibfield{author}{\bibinfo{person}{Ollie Johnston} {and} \bibinfo{person}{Frank Thomas}.} \bibinfo{year}{1981}\natexlab{}.
\newblock \bibinfo{booktitle}{\emph{The illusion of life: Disney Animation}}.
\newblock \bibinfo{publisher}{Disney Editions}.
\newblock


\bibitem[Joshi et~al\mbox{.}(2012)]%
        {Joshi:12:Cliplets}
\bibfield{author}{\bibinfo{person}{Neel Joshi}, \bibinfo{person}{Sisil Mehta}, \bibinfo{person}{Steven Drucker}, \bibinfo{person}{Eric Stollnitz}, \bibinfo{person}{Hugues Hoppe}, \bibinfo{person}{Matt Uyttendaele}, {and} \bibinfo{person}{Michael Cohen}.} \bibinfo{year}{2012}\natexlab{}.
\newblock \showarticletitle{Cliplets: juxtaposing still and dynamic imagery}. In \bibinfo{booktitle}{\emph{Proceedings of the 25th Annual ACM Symposium on User Interface Software and Technology}} (Cambridge, Massachusetts, USA) \emph{(\bibinfo{series}{UIST '12})}. \bibinfo{publisher}{Association for Computing Machinery}, \bibinfo{address}{New York, NY, USA}, \bibinfo{pages}{251–260}.
\newblock
\showISBNx{9781450315807}
\urldef\tempurl%
\url{https://doi.org/10.1145/2380116.2380149}
\showDOI{\tempurl}


\bibitem[Kazi et~al\mbox{.}(2014a)]%
        {Kazi:Draco}
\bibfield{author}{\bibinfo{person}{Rubaiat~Habib Kazi}, \bibinfo{person}{Fanny Chevalier}, \bibinfo{person}{Tovi Grossman}, \bibinfo{person}{Shengdong Zhao}, {and} \bibinfo{person}{George Fitzmaurice}.} \bibinfo{year}{2014}\natexlab{a}.
\newblock \showarticletitle{Draco: bringing life to illustrations with kinetic textures}. In \bibinfo{booktitle}{\emph{Proceedings of the SIGCHI Conference on Human Factors in Computing Systems}} (Toronto, Ontario, Canada) \emph{(\bibinfo{series}{CHI '14})}. \bibinfo{publisher}{Association for Computing Machinery}, \bibinfo{address}{New York, NY, USA}, \bibinfo{pages}{351–360}.
\newblock
\showISBNx{9781450324731}
\urldef\tempurl%
\url{https://doi.org/10.1145/2556288.2556987}
\showDOI{\tempurl}


\bibitem[Kazi et~al\mbox{.}(2014b)]%
        {Kazi:KineticTextures}
\bibfield{author}{\bibinfo{person}{Rubaiat~Habib Kazi}, \bibinfo{person}{Fanny Chevalier}, \bibinfo{person}{Tovi Grossman}, \bibinfo{person}{Shengdong Zhao}, {and} \bibinfo{person}{George Fitzmaurice}.} \bibinfo{year}{2014}\natexlab{b}.
\newblock \showarticletitle{Draco: Bringing Life to Illustrations with Kinetic Textures}. In \bibinfo{booktitle}{\emph{Proceedings of the SIGCHI Conference on Human Factors in Computing Systems}} (Toronto, Ontario, Canada) \emph{(\bibinfo{series}{CHI '14})}. \bibinfo{publisher}{Association for Computing Machinery}, \bibinfo{address}{New York, NY, USA}, \bibinfo{pages}{351–360}.
\newblock
\showISBNx{9781450324731}
\urldef\tempurl%
\url{https://doi.org/10.1145/2556288.2556987}
\showDOI{\tempurl}


\bibitem[Kazi et~al\mbox{.}(2016)]%
        {Kazi:2016:MotionAmplifiers}
\bibfield{author}{\bibinfo{person}{Rubaiat~Habib Kazi}, \bibinfo{person}{Tovi Grossman}, \bibinfo{person}{Nobuyuki Umetani}, {and} \bibinfo{person}{George Fitzmaurice}.} \bibinfo{year}{2016}\natexlab{}.
\newblock \showarticletitle{Motion Amplifiers: Sketching Dynamic Illustrations Using the Principles of 2D Animation}. In \bibinfo{booktitle}{\emph{Proceedings of the 2016 CHI Conference on Human Factors in Computing Systems}} (San Jose, California, USA) \emph{(\bibinfo{series}{CHI '16})}. \bibinfo{publisher}{Association for Computing Machinery}, \bibinfo{address}{New York, NY, USA}, \bibinfo{pages}{4599–4609}.
\newblock
\showISBNx{9781450333627}
\urldef\tempurl%
\url{https://doi.org/10.1145/2858036.2858386}
\showDOI{\tempurl}


\bibitem[Li et~al\mbox{.}(2023)]%
        {li2023chain}
\bibfield{author}{\bibinfo{person}{Chengshu Li}, \bibinfo{person}{Jacky Liang}, \bibinfo{person}{Andy Zeng}, \bibinfo{person}{Xinyun Chen}, \bibinfo{person}{Karol Hausman}, \bibinfo{person}{Dorsa Sadigh}, \bibinfo{person}{Sergey Levine}, \bibinfo{person}{Li Fei-Fei}, \bibinfo{person}{Fei Xia}, {and} \bibinfo{person}{Brian Ichter}.} \bibinfo{year}{2023}\natexlab{}.
\newblock \bibinfo{title}{Chain of Code: Reasoning with a Language Model-Augmented Code Emulator}.
\newblock
\newblock
\showeprint[arxiv]{2312.04474}~[cs.CL]


\bibitem[Li et~al\mbox{.}(2022)]%
        {alphacode}
\bibfield{author}{\bibinfo{person}{Yujia Li}, \bibinfo{person}{David Choi}, \bibinfo{person}{Junyoung Chung}, \bibinfo{person}{Nate Kushman}, \bibinfo{person}{Julian Schrittwieser}, \bibinfo{person}{Rémi Leblond}, \bibinfo{person}{Tom Eccles}, \bibinfo{person}{James Keeling}, \bibinfo{person}{Felix Gimeno}, \bibinfo{person}{Agustin Dal~Lago}, \bibinfo{person}{Thomas Hubert}, \bibinfo{person}{Peter Choy}, \bibinfo{person}{Cyprien de Masson~d’Autume}, \bibinfo{person}{Igor Babuschkin}, \bibinfo{person}{Xinyun Chen}, \bibinfo{person}{Po-Sen Huang}, \bibinfo{person}{Johannes Welbl}, \bibinfo{person}{Sven Gowal}, \bibinfo{person}{Alexey Cherepanov}, \bibinfo{person}{James Molloy}, \bibinfo{person}{Daniel~J. Mankowitz}, \bibinfo{person}{Esme Sutherland~Robson}, \bibinfo{person}{Pushmeet Kohli}, \bibinfo{person}{Nando de Freitas}, \bibinfo{person}{Koray Kavukcuoglu}, {and} \bibinfo{person}{Oriol Vinyals}.} \bibinfo{year}{2022}\natexlab{}.
\newblock \showarticletitle{Competition-level code generation with AlphaCode}.
\newblock \bibinfo{journal}{\emph{Science}} \bibinfo{volume}{378}, \bibinfo{number}{6624} (\bibinfo{date}{Dec.} \bibinfo{year}{2022}), \bibinfo{pages}{1092–1097}.
\newblock
\showISSN{1095-9203}
\urldef\tempurl%
\url{https://doi.org/10.1126/science.abq1158}
\showDOI{\tempurl}


\bibitem[Ma et~al\mbox{.}(2022)]%
        {Ma:Stylized3D}
\bibfield{author}{\bibinfo{person}{Jiaju Ma}, \bibinfo{person}{Li-Yi Wei}, {and} \bibinfo{person}{Rubaiat~Habib Kazi}.} \bibinfo{year}{2022}\natexlab{}.
\newblock \showarticletitle{A Layered Authoring Tool for Stylized 3D Animations}. In \bibinfo{booktitle}{\emph{Proceedings of the 2022 CHI Conference on Human Factors in Computing Systems}} (New Orleans, LA, USA) \emph{(\bibinfo{series}{CHI '22})}. \bibinfo{publisher}{Association for Computing Machinery}, \bibinfo{address}{New York, NY, USA}, Article \bibinfo{articleno}{383}, \bibinfo{numpages}{14}~pages.
\newblock
\showISBNx{9781450391573}
\urldef\tempurl%
\url{https://doi.org/10.1145/3491102.3501894}
\showDOI{\tempurl}


\bibitem[Nijkamp et~al\mbox{.}(2023)]%
        {codegen}
\bibfield{author}{\bibinfo{person}{Erik Nijkamp}, \bibinfo{person}{Bo Pang}, \bibinfo{person}{Hiroaki Hayashi}, \bibinfo{person}{Lifu Tu}, \bibinfo{person}{Huan Wang}, \bibinfo{person}{Yingbo Zhou}, \bibinfo{person}{Silvio Savarese}, {and} \bibinfo{person}{Caiming Xiong}.} \bibinfo{year}{2023}\natexlab{}.
\newblock \showarticletitle{CodeGen: An Open Large Language Model for Code with Multi-Turn Program Synthesis}.
\newblock \bibinfo{journal}{\emph{ICLR}} (\bibinfo{year}{2023}).
\newblock


\bibitem[Niklaus et~al\mbox{.}(2017)]%
        {Niklaus:2017:FrameInterpolation}
\bibfield{author}{\bibinfo{person}{Simon Niklaus}, \bibinfo{person}{Long Mai}, {and} \bibinfo{person}{Feng Liu}.} \bibinfo{year}{2017}\natexlab{}.
\newblock \bibinfo{title}{Video Frame Interpolation via Adaptive Convolution}.
\newblock
\newblock
\showeprint[arxiv]{1703.07514}~[cs.CV]
\urldef\tempurl%
\url{https://arxiv.org/abs/1703.07514}
\showURL{%
\tempurl}


\bibitem[Nye et~al\mbox{.}(2021)]%
        {nye2021scratchpads}
\bibfield{author}{\bibinfo{person}{Maxwell Nye}, \bibinfo{person}{Anders~Johan Andreassen}, \bibinfo{person}{Guy Gur-Ari}, \bibinfo{person}{Henryk Michalewski}, \bibinfo{person}{Jacob Austin}, \bibinfo{person}{David Bieber}, \bibinfo{person}{David Dohan}, \bibinfo{person}{Aitor Lewkowycz}, \bibinfo{person}{Maarten Bosma}, \bibinfo{person}{David Luan}, \bibinfo{person}{Charles Sutton}, {and} \bibinfo{person}{Augustus Odena}.} \bibinfo{year}{2021}\natexlab{}.
\newblock \bibinfo{title}{Show Your Work: Scratchpads for Intermediate Computation with Language Models}.
\newblock
\newblock
\showeprint[arxiv]{2112.00114}~[cs.LG]


\bibitem[Ralabate(2014)]%
        {ux_animation}
\bibfield{author}{\bibinfo{person}{Val Ralabate}.} \bibinfo{year}{2014}\natexlab{}.
\newblock \showarticletitle{UI Animation and UX: A Not-So-Secret Friendship}.
\newblock \bibinfo{journal}{\emph{A List Apart}} (\bibinfo{year}{2014}).
\newblock
\urldef\tempurl%
\url{https://alistapart.com/article/ui-animation-and-ux-a-not-so-secret-friendship/#section4}
\showURL{%
\tempurl}
\newblock
\shownote{Accessed: 2024-09-09}.


\bibitem[Ratajska et~al\mbox{.}(2020)]%
        {socialshapes}
\bibfield{author}{\bibinfo{person}{Adrianna Ratajska}, \bibinfo{person}{Matt~I Brown}, {and} \bibinfo{person}{Christopher~F Chabris}.} \bibinfo{year}{2020}\natexlab{}.
\newblock \showarticletitle{Attributing Social Meaning to Animated Shapes: A New Experimental Study of Apparent Behavior}.
\newblock \bibinfo{journal}{\emph{The American Journal of Psychology}} \bibinfo{volume}{133}, \bibinfo{number}{3} (\bibinfo{date}{10} \bibinfo{year}{2020}), \bibinfo{pages}{295--312}.
\newblock
\urldef\tempurl%
\url{https://doi.org/10.5406/amerjpsyc.133.3.0295}
\showDOI{\tempurl}


\bibitem[Rosenberg et~al\mbox{.}(2024)]%
        {Rosenberg:24:DrawTalking}
\bibfield{author}{\bibinfo{person}{Karl~Toby Rosenberg}, \bibinfo{person}{Rubaiat~Habib Kazi}, \bibinfo{person}{Li-Yi Wei}, \bibinfo{person}{Haijun Xia}, {and} \bibinfo{person}{Ken Perlin}.} \bibinfo{year}{2024}\natexlab{}.
\newblock \bibinfo{title}{DrawTalking: Building Interactive Worlds by Sketching and Speaking}.
\newblock
\newblock
\showeprint[arxiv]{2401.05631}~[cs.HC]
\urldef\tempurl%
\url{https://arxiv.org/abs/2401.05631}
\showURL{%
\tempurl}


\bibitem[Rozière et~al\mbox{.}(2024)]%
        {codellama}
\bibfield{author}{\bibinfo{person}{Baptiste Rozière}, \bibinfo{person}{Jonas Gehring}, \bibinfo{person}{Fabian Gloeckle}, \bibinfo{person}{Sten Sootla}, \bibinfo{person}{Itai Gat}, \bibinfo{person}{Xiaoqing~Ellen Tan}, \bibinfo{person}{Yossi Adi}, \bibinfo{person}{Jingyu Liu}, \bibinfo{person}{Romain Sauvestre}, \bibinfo{person}{Tal Remez}, \bibinfo{person}{Jérémy Rapin}, \bibinfo{person}{Artyom Kozhevnikov}, \bibinfo{person}{Ivan Evtimov}, \bibinfo{person}{Joanna Bitton}, \bibinfo{person}{Manish Bhatt}, \bibinfo{person}{Cristian~Canton Ferrer}, \bibinfo{person}{Aaron Grattafiori}, \bibinfo{person}{Wenhan Xiong}, \bibinfo{person}{Alexandre Défossez}, \bibinfo{person}{Jade Copet}, \bibinfo{person}{Faisal Azhar}, \bibinfo{person}{Hugo Touvron}, \bibinfo{person}{Louis Martin}, \bibinfo{person}{Nicolas Usunier}, \bibinfo{person}{Thomas Scialom}, {and} \bibinfo{person}{Gabriel Synnaeve}.} \bibinfo{year}{2024}\natexlab{}.
\newblock \bibinfo{title}{Code Llama: Open Foundation Models for Code}.
\newblock
\newblock
\showeprint[arxiv]{2308.12950}~[cs.CL]


\bibitem[Saquib et~al\mbox{.}(2019)]%
        {Saquib:19:Interactive}
\bibfield{author}{\bibinfo{person}{Nazmus Saquib}, \bibinfo{person}{Rubaiat~Habib Kazi}, \bibinfo{person}{Li-Yi Wei}, {and} \bibinfo{person}{Wilmot Li}.} \bibinfo{year}{2019}\natexlab{}.
\newblock \showarticletitle{Interactive Body-Driven Graphics for Augmented Video Performance}. In \bibinfo{booktitle}{\emph{Proceedings of the 2019 CHI Conference on Human Factors in Computing Systems}} (Glasgow, Scotland Uk) \emph{(\bibinfo{series}{CHI '19})}. \bibinfo{publisher}{Association for Computing Machinery}, \bibinfo{address}{New York, NY, USA}, \bibinfo{pages}{1–12}.
\newblock
\showISBNx{9781450359702}
\urldef\tempurl%
\url{https://doi.org/10.1145/3290605.3300852}
\showDOI{\tempurl}


\bibitem[Si et~al\mbox{.}(2024)]%
        {design2code}
\bibfield{author}{\bibinfo{person}{Chenglei Si}, \bibinfo{person}{Yanzhe Zhang}, \bibinfo{person}{Zhengyuan Yang}, \bibinfo{person}{Ruibo Liu}, {and} \bibinfo{person}{Diyi Yang}.} \bibinfo{year}{2024}\natexlab{}.
\newblock \bibinfo{title}{Design2Code: How Far Are We From Automating Front-End Engineering?}
\newblock
\newblock
\showeprint[arxiv]{2403.03163}~[cs.CL]


\bibitem[Sur\'is et~al\mbox{.}(2023)]%
        {vipergpt}
\bibfield{author}{\bibinfo{person}{D\'idac Sur\'is}, \bibinfo{person}{Sachit Menon}, {and} \bibinfo{person}{Carl Vondrick}.} \bibinfo{year}{2023}\natexlab{}.
\newblock \showarticletitle{ViperGPT: Visual Inference via Python Execution for Reasoning}.
\newblock \bibinfo{journal}{\emph{Proceedings of IEEE International Conference on Computer Vision (ICCV)}} (\bibinfo{year}{2023}).
\newblock


\bibitem[Team(2023)]%
        {gemini}
\bibfield{author}{\bibinfo{person}{Gemini Team}.} \bibinfo{year}{2023}\natexlab{}.
\newblock \bibinfo{title}{Gemini: A Family of Highly Capable Multimodal Models}.
\newblock
\newblock
\showeprint[arxiv]{2312.11805}~[cs.CL]


\bibitem[Tseng et~al\mbox{.}(2024)]%
        {Tseng:24:Keyframer}
\bibfield{author}{\bibinfo{person}{Tiffany Tseng}, \bibinfo{person}{Ruijia Cheng}, {and} \bibinfo{person}{Jeffrey Nichols}.} \bibinfo{year}{2024}\natexlab{}.
\newblock \bibinfo{title}{Keyframer: Empowering Animation Design using Large Language Models}.
\newblock
\newblock
\showeprint[arxiv]{2402.06071}~[cs.HC]


\bibitem[Willett et~al\mbox{.}(2018)]%
        {Willett:18:MixedInitiative}
\bibfield{author}{\bibinfo{person}{Nora~S. Willett}, \bibinfo{person}{Rubaiat~Habib Kazi}, \bibinfo{person}{Michael Chen}, \bibinfo{person}{George Fitzmaurice}, \bibinfo{person}{Adam Finkelstein}, {and} \bibinfo{person}{Tovi Grossman}.} \bibinfo{year}{2018}\natexlab{}.
\newblock \showarticletitle{A Mixed-Initiative Interface for Animating Static Pictures}. In \bibinfo{booktitle}{\emph{Proceedings of the 31st Annual ACM Symposium on User Interface Software and Technology}} (Berlin, Germany) \emph{(\bibinfo{series}{UIST '18})}. \bibinfo{publisher}{Association for Computing Machinery}, \bibinfo{address}{New York, NY, USA}, \bibinfo{pages}{649–661}.
\newblock
\showISBNx{9781450359481}
\urldef\tempurl%
\url{https://doi.org/10.1145/3242587.3242612}
\showDOI{\tempurl}


\bibitem[Zhang et~al\mbox{.}(2023)]%
        {Zhang:2023:Motion}
\bibfield{author}{\bibinfo{person}{Sharon Zhang}, \bibinfo{person}{Jiaju Ma}, \bibinfo{person}{Jiajun Wu}, \bibinfo{person}{Daniel Ritchie}, {and} \bibinfo{person}{Maneesh Agrawala}.} \bibinfo{year}{2023}\natexlab{}.
\newblock \showarticletitle{Editing Motion Graphics Video via Motion Vectorization and Transformation}.
\newblock \bibinfo{journal}{\emph{ACM Transactions on Graphics}} \bibinfo{volume}{42}, \bibinfo{number}{6} (\bibinfo{date}{Dec.} \bibinfo{year}{2023}), \bibinfo{pages}{1–13}.
\newblock
\showISSN{1557-7368}
\urldef\tempurl%
\url{https://doi.org/10.1145/3618316}
\showDOI{\tempurl}


\end{thebibliography}

\end{document}

\section*{Supplemental Material}

\appendix

\subsection*{Comparison Study Metric Rubrics}

\textbf{Relevance Rubric.}

Relevance - does the animation match the subject of the logo?

\begin{itemize}

\item 1,  Not relevant: The animation does not match the subject of the logo
\item 2, Minimally relevant. The animation minimally matches the subject of the logo
\item 3, Moderately relevant: The animation moderately matches the subject of the logo
\item 4, Good relevance: The animation matches the subject of the logo
\item 5, Strongly relevant: Animation strongly matches the subject of the logo

\end{itemize}

\textbf{Sequencing - overall, how well is the animation sequenced in terms of the coordination and pacing of elements? }

\begin{itemize} 

\item 1, Very poor sequencing. Coordinatinon and pacing has significant issues (e.g. elements come in out of order, groups are illogical, pacing is unacceptable)
\item 2, Poor sequencing. Sequencing is minimal or lacking in coordination and pacing across elements
\item 3, Acceptable sequencing. Sequencing shows adequate coordination and timing across elements
\item 4,  Good sequencing. Sequencing shows good coordination and pacing
\item 5, Excellent sequencing.  Animation has excellent coordination across elements and pacing (e.g. logical groups, good pacing, proper ordering, etc.)

\end{itemize}
\textbf{Execution Quality - how well was the animation executed, where there any flaws? Flaws can be things like: elements don't make it to their final state, overanimation, odd cropping)}

\begin{itemize}
\item 1,  Very poor execution quality. Animations have critical flaws
\item 2,  Poor execution quality. Animations has multiple flaws
\item 3,  Acceptable Execution Quality. Animation is passable with noncritical flaws
\item 4,  Good Execution Quality - Animation is executed well with minimal flaws 
\item 5, Excellent Execution Quality - Animation is well-executed with no flaws

\end{itemize}